\documentclass[twocolumn]{aastex63}
\usepackage{graphicx}
\usepackage{amsmath}
\usepackage{amssymb}
\usepackage{apjfonts}

\newcommand{\cmc}{\,{\rm cm$^{-3}$}\,}
\newcommand{\kms}{\,{\rm km\,s$^{-1}$}\,} 
\newcommand{\kmsmpc}{\,{\rm km\,s$^{-1}$\,Mpc$^{-1}$}\,}

\newcommand{\etal}{{ et~al.~}}

\newcommand{\ergscm}{\,{\rm erg\,s$^{-1}$\,cm$^{-2}$}\,}
\newcommand{\Ms}{M_\odot}

\newcommand{\Zs}{Z_\odot}

\newcommand{\as}{^{\prime\prime}}

\shorttitle{A Major Galaxy Group Merger at Apogee}
\shortauthors{Machacek et al.}

\begin{document}


\title{A Chandra Study of the NGC\,7618/UGC\,12491 Major Group Merger at Apogee: Multiple Cold Fronts, Boxy Wings, Filaments and Arc-Shaped Slingshot Tails}

\author{M. E. Machacek}
\affiliation{Center for Astrophysics | Harvard \& Smithsonian, 60 Garden Street, Cambridge, MA 02138 USA}
\author{C. Jones}
\affiliation{Center for Astrophysics | Harvard \& Smithsonian, 60 Garden Street, Cambridge, MA 02138 USA}
\author{R. P. Kraft}
\affiliation{Center for Astrophysics | Harvard \& Smithsonian, 60 Garden Street, Cambridge, MA 02138 USA}
\author{ W. R. Forman}
\affiliation{Center for Astrophysics | Harvard \& Smithsonian, 60 Garden Street, Cambridge, MA 02138 USA}
\author{E. Roediger}
\affiliation{E. A. Milne Centre for Astrophysics, Department of Physics and Mathematics, University of Hull, Hull HU6 7RX, UK}
\author{A. Sheardown}
\affiliation{E. A. Milne Centre for Astrophysics, Department of Physics and Mathematics, University of Hull, Hull HU6 7RX, UK}
\author {J. Wan}
\affiliation{Division of Physics, Mathematics and Astronomy, California Institute of Technology, 1200 E California Blvd, Pasadena, CA 91125 USA}

\email{mmachacek@cfa.harvard.edu}

\begin{abstract}
\noindent Analyses of major group mergers are key to 
understanding the evolution of large-scale structure in the Universe 
and the microphysical properties of the hot gas in these systems.
We present imaging and spectral analyses of deep {\em Chandra} observations
of hot gas structures formed in the major merger of the NGC\,7618 and
UGC\,12491 galaxy groups and compare the observed hot gas morphology, 
temperatures and abundances with recent simulations.  The morphology of 
the observed  multiple cold front edges and boxy wings are consistent 
with those expected to be formed by Kelvin-Helmholtz instabilities and 
gas sloshing in inviscid gas. The arc-shaped slingshot tail morphologies 
seen in each galaxy suggest that the dominant galaxies are near their 
orbital apogee after having experienced at least one core passage at a
large impact parameter. 

\end{abstract}



\section{Introduction}
\label{sec:intro}

Galaxy groups and clusters grow through the merger of smaller
systems. X-ray observations that measure the temperature, density and
entropy of the hot gas in merging galaxy groups and clusters show 
a wide range of complex and varied morphologies. Among these are
X-ray surface brightness discontinuities (cold
fronts or shocks), sweeping spiral features caused by  
bulk motions of the gas relative to the disturbed dark matter gravitational
potential (sloshing), extended, sometimes arching tails, either stripped or
partially stripped from the merging systems, and boxy, scalloped
or filamentary features characteristic of hydrodynamical instabilities,  
such as Kelvin-Helmholtz Instabilities (KHI). For examples of previous
studies of these morphologies in galaxy group and cluster mergers, see, e.g., 
the review by Markevitch \& Vikhlinin 2007 and references therein. 
Comparisons of deep {\em Chandra} X-ray  observations of complex gas
features in nearby mergers with high resolution numerical simulations matched 
to the merging system allow  us to investigate the orbital dynamics,
configuration and age of the merging partners, while also
providing a powerful tool to constrain the
microphysical properties (thermal conduction, magnetic field,
viscosity), chemical abundance, and dynamical state of the surrounding
intracluster medium (ICM) (ZuHone \etal 2011, Roediger \etal 2012b,
Roediger \etal 2013, ZuHone \etal 2013, Roediger \etal 2015a, 2015b,  
Kraft \etal 2017, Su \etal 2017a, 2017b). Such studies 
give valuable insights into the role mergers play in the evolution
and metallicity of the ICM. 

\begin{figure*}[htb!]
\begin{center}
\includegraphics[width=5in]{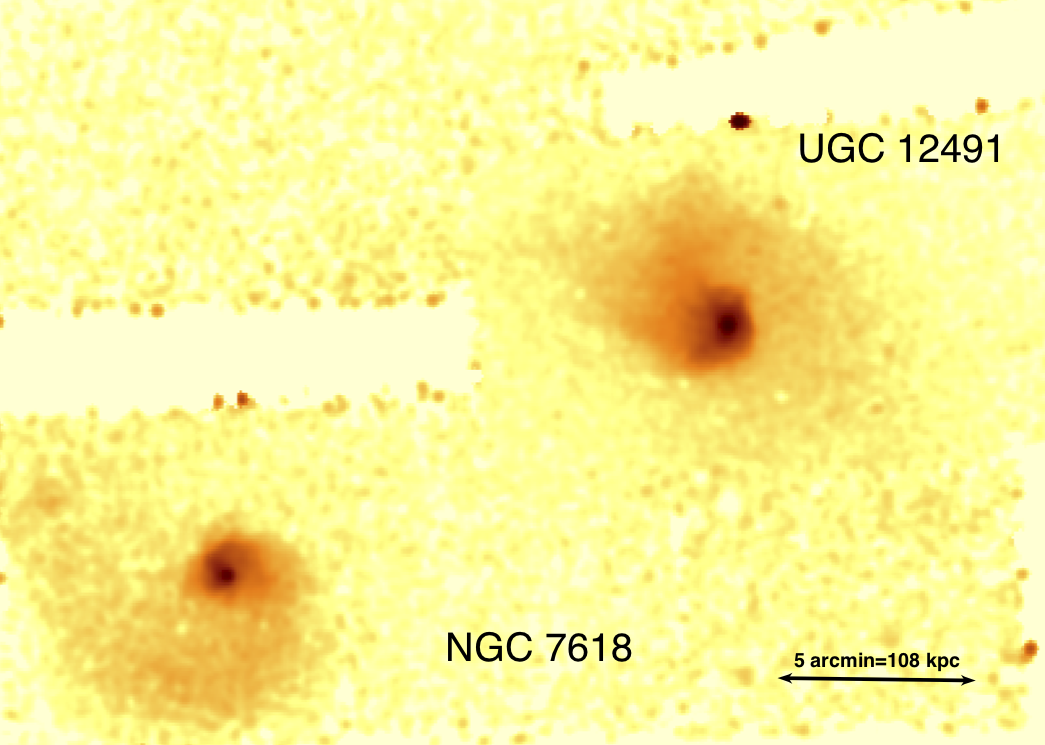}
\includegraphics[width=5in]{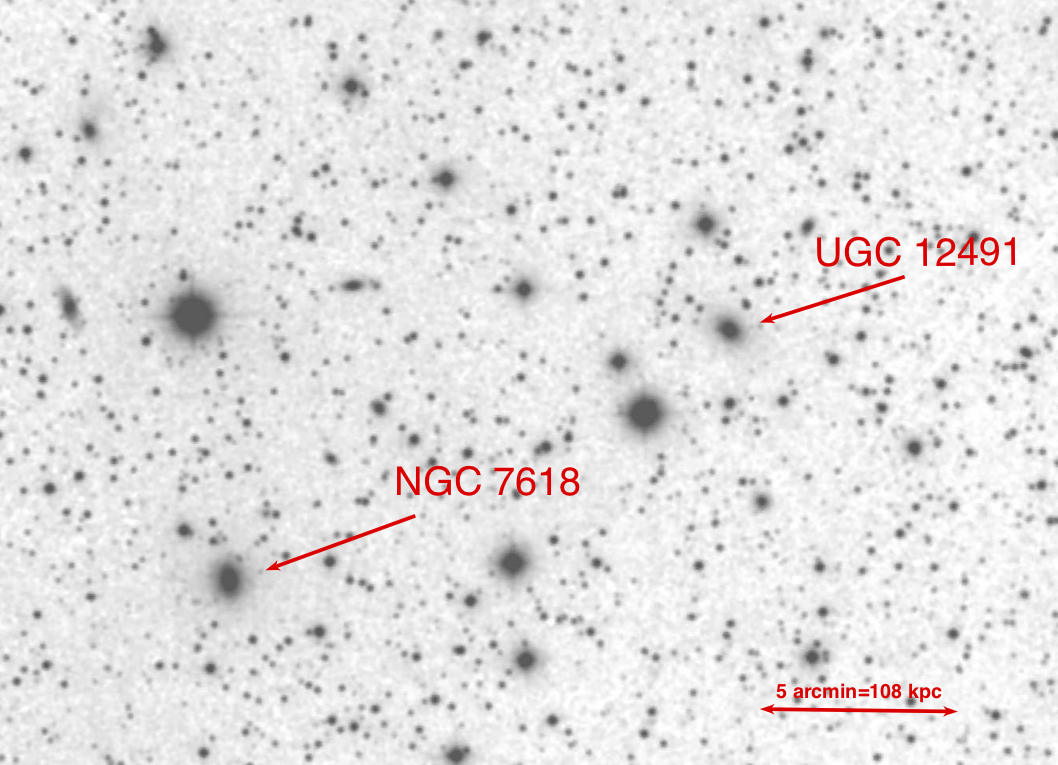}
\caption{((upper panel) Exposure-corrected, background-subtracted, 
  co-added  {\em Chandra} X-ray image in the $0.5-2.0$\,keV energy 
  band of the hot gas associated with the NGC\,7618 (lower left) and
  the UGC\,12491 (upper right) galaxy groups.  
  Point sources have been excluded. The data have been binned by $8 \times 8$
  {\it Chandra} pixels, 
  such that in the image $1\,{\rm pixel} = 3.936\as \times  3.936\as$, 
  and smoothed with a $3 \sigma$ Gaussian kernel to highlight the 
   faint emission in the tails. North is up and east is to the left. 
(lower panel) DSS optical image of NGC\,7618 (lower left) and UGC\,12491
(upper right) matched to the {\em Chandra} X-ray image in WCS coordinates.}
\label{fig:merge}
\end{center}
\end{figure*}
 
Over the past two decades, the {\em Chandra} and {\em XMM-Newton} X-ray
Observatories have allowed detailed studies of the properties of hot
gas in many merging galaxy clusters. These studies have
focused primarily on energetic mergers in bright massive clusters, e.g. shocks
and merger cold fronts in the Bullet Cluster (1E0657-56) 
(Markevitch \etal 2002b), 
Abell 521 (Bourdin \etal 2013), and Abell 665 (Dasadia \etal 2016),
long sloshing spirals in Abell 2029 (Paterno-Mahler \etal 2013), and both
shocks and sloshing features in the subcluster merger in  RXJ 1347.5-1145 
(Johnson \etal 2012; Kreisch \etal 2016) and Abell 115 (Forman \etal 2010;
 Botteon \etal 2016; Hallman \etal 2018).  Most deep {\em Chandra} 
and {\em XMM-Newton} X-ray observations of galaxy-group-scale mergers
focus on minor mergers, 
i.e the accretion and gas stripping of the galaxy or galaxy group as it
infalls through the hot gas atmosphere of a more massive cluster. 
When compared to numerical simulations, 
these studies investigate the physics of gas stripping (ram pressure, KHI, 
turbulence, sloshing) and its impact on gas temperatures and gas mixing in
the group. 
See, for example, NGC\,1404 in Fornax 
(Machacek \etal 2005; Scharf \etal 2005; Su \etal 2017a, 2017b; 
Sheardown \etal 2018), NGC\,4552 (Machacek \etal 2006; 
 Roediger \etal 2015a, 2015b; Kraft \etal 2017), NGC\,4472 (Kraft \etal
2011; Sheardown \etal 2019), M86 (Randall \etal 2008) and 
M60 (Wood \etal 2017; Sheardown  \etal 2019), all in the Virgo cluster, 
LEDA in Hydra A
(De Grandi \etal 2016; Sheardown \etal 2019), and the G3-G5 group in 
Abell 2142 (Eckert \etal 2017). However, 
these examples do not probe the earliest stages of structure
formation, when low mass groups first begin to merge.  

Compared to clusters, mergers within and between galaxy groups 
are more difficult to study because of their lower X-ray luminosities. 
Previous  studies have focused on the identification of sloshing features 
in the dominant group galaxy caused by a recent off-axis encounter
with another galaxy in the group, such as in NGC\,5846 (Machacek
\etal 2011), NGC\,5044 (David \etal 2009; Gastaldello \etal 2009;
O'Sullivan \etal 2014), and IC\,1860 (Gastaldello \etal 2013). 
Mergers between galaxy groups of equal or comparable size (major
mergers) are not as well studied. Note, however, the disturbed X-ray
morphologies found in the merging NGC\,6868 and NGC\,6861 groups
(Machacek \etal 2010), the merging cold fronts in the NGC\,7619 and
NGC\,7626 groups in Pegasus (Randall \etal 2009), and the more 
recent deep {\em Chandra} observations 
of extremely violent (Mach $>2$) galaxy group mergers concurrent 
with active galactic nuclei (AGN) activity in RXJ0751.3+5012
(Russell \etal 2014) and NGC\,6338 (O'Sullivan \etal 2019). 

In this paper, we discuss the interaction of the \object{NGC 7618} and
\object{UGC 12491} galaxy groups.  
 At a luminosity distance of $75$\,Mpc, angular separation of only 
$14.1^{\prime}$ and radial velocity difference $\delta v_r = 17$\kms
(Huchra \etal 1999),  NGC\,7618 and  UGC\,12491 might appear  
to be natural candidates for an interacting system. However,
due to their low Galactic latitudes ($\sim -17^\circ$) and apparently 
undisturbed stellar morphologies (see the lower panel of 
Figure \ref{fig:merge}), 
early optical and infrared studies of these nearby elliptical galaxies
misidentified them as isolated galaxies (Cobert \etal 2001).
NGC\,7618 hosts an ultra-steep spectrum radio source
($\gamma (325{\rm  MHz}- 1.4{\rm GHz})=-1.42$) at its optical center, 
suggesting the presence of an AGN (De\,Brueck \etal 2000).
This is further evidenced 
by the ROSAT observation  of a bright X-ray point
source (1RXSJ231947.4+4) coincident with NGC\,7618's center 
(Voges \etal 1999). UGC\,12491 also hosts a weak, compact radio source 
at its optical center consistent with an AGN (Condon \etal 2002). 
However, extended radio structures have not been observed in 
either galaxy.

Using X-ray observations of NGC\,7618 and UGC\,12491 taken with the 
{\em Advanced Satellite for Cosmology and Astrophysics} Gas Imaging 
Spectrometer (ASCA GIS), Kraft \etal (2006) found   highly asymmetric, 
$\sim 1$\,keV X-ray 
emission extending to radii of $150-200$\,kpc from each 
galaxy, far greater than expected for the hot gas halo of an
elliptical galaxy alone. The hot gas temperatures 
and total X-ray luminosities around each
galaxy were more typical of that expected for poor galaxy clusters or 
fossil groups. In their analysis of the ASCA
GIS images, Kraft \etal (2006) also found that the galaxy groups were 
embedded in faint, diffuse gas with a temperature of $\sim 2$\,keV, 
albeit with large uncertainties, suggesting that the
groups  may be bound within a more massive dark matter halo. 
From a 1999 December 10  {\em Chandra} observation of NGC\,7618 (ObsId
802) with a useful exposure of $\sim 8.4$\,ks, Kraft \etal (2006) 
identified a sharp surface brightness discontinuity to the north and a
faint tail to the south of NGC\,7618.
They concluded that the observed features were merger
induced, either through the major merger of two roughly equal mass
galaxy groups or through the infall of each group into a more massive 
dark matter potential. This merging hypothesis was strengthened by Crook \etal
(2007, 2008),  who used a variable-linking-length percolation group finding 
algorithm (Huchra \& Geller 1982) on galaxy data from the Two Micron
All Sky Redshift Survey (Huchra \etal 2012) to search for  nearby 
galaxy groups down to low Galactic latitudes. They found  
NGC\,7618, UGC\,12491 and eight additional galaxies to be members of a galaxy
 group (HCD\,1239) with group virial mass, virial radius and line-of-sight 
velocity dispersion of $3.96 \times 10^{13}\Ms$, $1.1$\,Mpc, and $181.7$\kms,
respectively. However, due to the low galaxy number density,
they could not identify substructure within the HCD\,1239 group. 

In 2007 NGC\,7618 and UGC\,12491 were each observed again with 
{\em Chandra} ACIS-S for effective exposures of $33$\,ks and
$31.7$\,ks, respectively. The analysis of the X-ray data by Roediger
\etal (2012a) confirmed the hypothesis of an off-axis merger of two
roughly equal mass galaxy groups.  They found the hot gas 
in and around each galaxy to be highly disturbed, with arc-like cold fronts, 
long curved tails, and wings and boxy features suggestive of 
KHI, consistent with features found in numerical simulations of more
massive galaxy cluster mergers (ZuHone \etal 2010; Roediger \etal 2012b). 
The disturbed X-ray gas morphologies, observed in each
galaxy group by Roediger \etal (2012a) and studied in more detail in this
work (e.g., see Fig. \ref{fig:merge}), 
can be used to probe the tidal and gas hydrodynamical forces at work in the 
merger without the added complication of strong AGN feedback. This  makes  
the merger of NGC\,7618/UGC\,12491 an ideal laboratory to measure the
properties of merger-induced hydrodynamical instabilities, such as
KHI. Thus, the NGC\,7618/UGC\,12491 merger is one of the best examples 
of a major merger of two, cool ($\sim 1$\,keV)  equal mass galaxy groups 
in the local universe.

Based on the small radial velocity difference and well-defined cold
fronts, previous work interpreted the merger of NGC\,7618 and
UGC\,12491 to be occurring in the 
plane of the sky (Kraft \etal 2006, Roediger \etal 2012).  
However, recent simulations suggest an alternative hypothesis, in particular, 
that the dominant group galaxies are  near the 
apogee of their orbits, with a  viewing angle for the merger 
$\sim 45^\circ$ to the LOS (Sheardown \etal 2019). At
apogee, the groups would have little motion relative to each other. Thus an
observation along any LOS  would produce an almost zero
 radial velocity difference. The inclination of the merger plane would, 
however, impact other characteristics, e.g. how tightly wound the spiral 
tails appear. 
 
In this work we combine the archival {\it Chandra} observations reported above
with more recent {\it Chandra} observations for total useful  exposures of  
$50.4$\,ks for NGC\,7618 and $147$\,ks ($98.7$\,ks) for imaging
(spectral) analyses of UGC\,12491, respectively.  In Figure
\ref{fig:merge} we show the mosaic of background-subtracted, 
exposure-corrected, co-added $0.5-2$\,keV images of NGC\,7618 and 
UGC\,12491, binned by $8$ {\it Chandra} pixels, to capture both merging 
galaxy groups and highlight faint X-ray features. Point sources, other 
than the galaxies' central AGNs, have been excluded to focus  
on the properties of the diffuse gas. 
The richness and complexity of the observed gas features within and around 
each dominant group galaxy are stunning. Using imaging and spectral analyses 
of these data, we show that the measured density and temperature
structure 
of the observed hot gas strongly suggest the latter interpretation is correct.

This paper is organized as follows: 
In \S\ref{sec:obs} we describe the data reduction and cleaning of the 
 {\it Chandra} observations used in our analysis.  In 
 \S\ref{sec:analysis} we use background-subtracted, exposure-corrected, 
co-added images of the merger to identify the X-ray gas 
morphologies in and  around each dominant galaxy. For each galaxy we 
characterize these features using simple gas density models to fit the
observed X-ray surface brightness profiles across the regions of interest,
and measure the corresponding temperatures and metal abundances of the
hot gas in the galaxies and in the surrounding intra-group gas (IGM).
In \S\ref{sec:discuss} we compare our results to the gas
morphologies found in simulations of equal mass mergers 
and expected from models of gas stripping via Kelvin Helmholtz 
Instabilities (KHI)
to determine the stage of the merger and microphysical properties of
the galaxy and intra-group gas. We summarize our results in 
\S\ref{sec:conclude}. All coordinates are J2000. Assuming a Hubble
constant of $H_0=69$\kmsmpc in a flat Lambda cold dark matter
cosmology ($\Omega_m=0.3$), and the luminosity distance to NGC\,7618 of
$75$\,Mpc, $1" = 0.36$\,kpc (Wright 2006). 


\section{Observations and data reduction}
\label{sec:obs}

In Table \ref{tab:ChandraObsID} we list the {\it Chandra} observations
used in this analysis. All of the observations use the Advanced CCD
Imaging Spectrometer array (ACIS) in VFAINT mode  with the
 back-illuminated chip S3 at the 
aimpoint. Three observations (ObsIDs 7896, 16015, and 17412) place 
UGC\,12491 near the aimpoint on chip S3,  while two observations 
(ObsIDs 7985 and 16014) place NGC\,7618 at the aimpoint.

Since source emission fills the S3 CCD for each observation, we
cannot use local background subtraction for our data. Instead we
use source free (blank sky) background data sets available in the
CIAO Calibration Data Base (CALDB) appropriate for the ACIS CCD
and date of observation. However, these blank sky data sets were
taken at high Galactic latitude, while the NGC\,7618 and
UGC\,12491 galaxy groups are at relatively low ($\sim -17^{\circ}$)
Galactic latitude such that additional backgrounds not accounted
for in the standard blank sky background subtraction procedure
may be a concern. To investigate this possibility, we also
analyze X-ray emission on the ACIS CCD S1 for ObsIDs 17412 and
16015. ACIS CCD S1 is also back-illuminated and was on in our
observations. For our observations, the orientation of the
ACIS detector places the center of the S1 CCD $\sim 16^{\prime}$
west of the S3 aimpoint. For observations with  NGC\,7618
at the S3 aimpoint, this is in the direction of UGC\,12491
and, as seen in Figure \ref{fig:merge}, positions S1 near (in
projection) UGC\,12491 where IGM X-ray emission from the
merging system is expected to be significant. This makes
the S1 data in these observations unsuitable for
for the study of additional backgrounds. In contrast, for
ObsIDs 17412 and 16015 with UGC\,12491 at their respective
aimpoints, the center of the S1 CCD lies (in projection)
$\sim 16^{\prime}$ west of UGC\,12491 in a direction away
from NGC\,7618, where the contribution to the X-ray
emission from the hot gas halo in the merger is
expected to be small.

All data were 
reprocessed using standard CIAO 4.9 tools to apply the most current 
calibrations and instrument corrections, including corrections for the 
charge transfer inefficiency on the ACIS CCDs, the time-dependent 
build-up of contaminant on the optical filter, and the secular drift 
of the average pulse-height amplitude for photons of fixed energy
(tgain). In addition to the standard rejection of events with bad 
array patterns (grades $1$, $5$, and $7$), events flagged by VFAINT 
mode as having excess counts in border pixels were also excluded in 
order to maximize signal-to-noise at energies below $2$\,keV, where 
most of the source emission is observed.

After point sources were excluded, the data were cleaned using the 
$2.5-7$\,keV energy band on CCD S3 and 
the  $2.5-6$\,kev energy band on CCD S1
to be consistent with the cleaning 
algorithm used for blank sky backgrounds for the S3
and S1 CCDs, respectively, and date of observation. 
 Light curves were extracted for each
 observation to identify large flares. For observations 7895, 7896,
 16015 and 17412, periods of anomalously high and low count rates were excluded 
using CIAO tools {\it deflare} and {\it lc\_clean} to compute the mean 
count rate using a $3\sigma$ algorithm and then excise time periods 
where the observed count rate deviated by more than $20\%$ from that
mean. Observation 16014 was contaminated by a class X1.6 solar coronal 
mass ejection (CME) $~20$\,ks after the start of the observation. 
Examination of the light curve showed significant post-flare
{\em ringing} of the CCD throughout the rest of the observation, such 
that we excluded all post flare data from that observation. This 
resulted in an effective exposure for ObsID 16014 of
$17.4$\,ks. Although the light curve for ObsID 16015, taken $7$ days 
prior to the major solar CME, showed no 
obvious flaring, the mean cleaned count rate was $77\%$ higher than 
the mean cleaned count rate for ObsID 17412   
taken $10$ days later with the same aimpoint. Such a rapid and large 
change  in the mean count rate after standard cleaning  
is symptomatic of low-level flare contamination throughout
ObsID 16015.

We used the CIAO tool {\em blanksky} to construct background
files for imaging and 
spectral analysis from the source-free (blank sky) background
data sets available in the CIAO Calibration Data Base (CALDB)
appropriate for the ACIS CCD and date of observation, and reproject
them onto the observation. 
Identical cleaning was applied to both the blank-sky
background and source data. The normalization of the blank-sky
background to each source observation was fine-tuned by matching the 
blank-sky background count rates to the respective source count rates 
in the $9 - 12$\,keV energy band, where particle backgrounds dominate.

 Ciao tool {\em blanksky\_image} was used to create source,
  blank sky, and blank-sky-subtracted images for the S1 data in the
  flare-free data set ObsId 17412 and the flared data set
  ObsId 16015.   After excluding
  point sources, we use the ObsId 17412 S1 source, blank-sky background and
  background-subtracted images in the $0.5-2$\,keV band of interest to
  study the importance of any residual background. We find excess
  emission in the $0.5-2$\,keV band is small ( $\sim 15\%$)
  relative to the blank-sky background and is uniform over the
  chip. This should have a negligible effect on the results of our
  analysis.

  We then use the $0.5-2$\,keV blank-sky-subtracted
  image of the S1 chip for ObsId 16015 to study the spatial
  distribution of the low level flare in this observation. We find
  that the excess emission shows no significant structure across the
  chip. Since this background due to soft (low level) flaring is found
  to be uniform across the ACIS back-lluminated CCDs in ObsId 16015,
  it  will not distort the imaging results for the 
 sloshing and hydrodynamic-instability-induced  features of interest 
 in this study. We thus include it in our imaging analysis.
 However, the spectral shape of such low-level flaring 
is different from that of the blank-sky backgrounds, so it cannot be 
eliminated from the spectral analysis by a simple renormalization 
of the blank-sky backgrounds at high energies. The spectral properties of
such soft flares adversely affect measurements of the spectral properties of 
interest for the diffuse gas in the merging galaxies.  APEC model fits to the 
solar flare contaminated data in ObsID 16015 yield anomalously high 
temperatures, inconsistent with the gas temperature measured from APEC 
model spectral fits to the other observations (ObsID 7896 and ObsID 17412) 
of UGC12491. Note that Markevitch (2002a) found a  similar anomalous 
increase in the fitted gas temperatures in Chandra observations of 
Abell 1835, taken using the ACIS back-illuminated CCD S3, due to 
contamination by soft flares. Thus we do not use data from ObsID 16015 
in our spectral analysis of UGC\,12491.

After cleaning, 
we found the  combined effective exposure for the NGC\,7618 group to
be $50,434$\,s.  The combined useful  exposure for the UGC\,12491 group is  
$147,128$\,s for imaging and $98,706$\,s for spectral analyses (see Table
\ref{tab:ChandraObsID}).
Background-subtracted source images from CCD S3 for each observation,
constructed using the CIAO tool 
{\em blanksky\_image},  were co-added and exposure corrected to create
mosaics for use in our imaging analysis. Point sources were excluded
from these mosaics to focus on the properties of the X-ray emission
from  the  diffuse hot gas.  

\begin{deluxetable}{ccccc}
\tablewidth{0pc}
\tablecaption{{{\it Chandra} Observations of UGC\,12491 and NGC\,7618}
\label{tab:ChandraObsID}}
\tablehead{\colhead{ObsID} & \colhead{Date} & \colhead{Exposure} 
& \colhead{Cleaned Exposure} \\
& & (s) & (s) }
\startdata
UGC\,12491 & & & \\
7896    &2007 Sep 3  &$33613$  &$31728$ \\
17412   &2014 Sep 13 &$68005$  &$66978$ \\
16015   &2014 Sep 3  &$48425$  &$48422$  \\
NGC\,7618 & & & \\
7895    &2007 Sep 8  & $34059$ & $33033$ \\
16014   &2014 Sep 10 &$119424$ & $17401$ \\
\enddata
\tablecomments{ObsID 16015 is excluded from the  spectral analyses of
  the UGC\,12491 group due to low level flare contamination that could
not be isolated and removed  from the data using standard techniques. }
\end{deluxetable}

Spectra for the regions of interest were extracted for identical source
and blank-sky background regions for each flare-cleaned observation
using the CIAO tool {\em  specextract}.
All resolved point sources including the central AGN for each
  galaxy were excluded from these spectral regions.
The background-subtracted spectra for each region
were fit jointly using the HEASOFT spectral fitting package 
{\em XSpec\,12.10.0e} (Arnaud 1996) with solar abundance tables from
Anders \& Grevesse (1989). Unless otherwise indicated, quoted
uncertainties are $90\%$ CL.

\section{Anatomy of a Merger: KHI Wings, Sloshing Edges 
and Tails}
\label{sec:analysis}

In sections \S\ref{sec:ugcanalysisintro} and
\S\ref{sec:ngcanalysisintro}, we use  high-resolution
$0.5-2$\,keV {\it  Chandra} X-ray images of UGC\,12491 and 
NGC\,7618, the dominant galaxy in each group, to characterize
the gas morphology in and around these galaxies. For 
each dominant group galaxy, we
use X-ray surface brightness profiles to identify surface brightness 
discontinuities (edges) and to study their tails. We parameterize the 
gas morphology by integrating simple gas density models along
the LOS to reproduce the observed surface brightness
profiles. We measure the hot gas temperatures and abundances along
these profiles, in the X-ray detected wings, tails and surrounding IGM, 
and calculate effective gas density, temperature and pressure ratios 
across the X-ray surface brightness edges.
  
\begin{figure*}
\begin{center}
\includegraphics[width=5in]{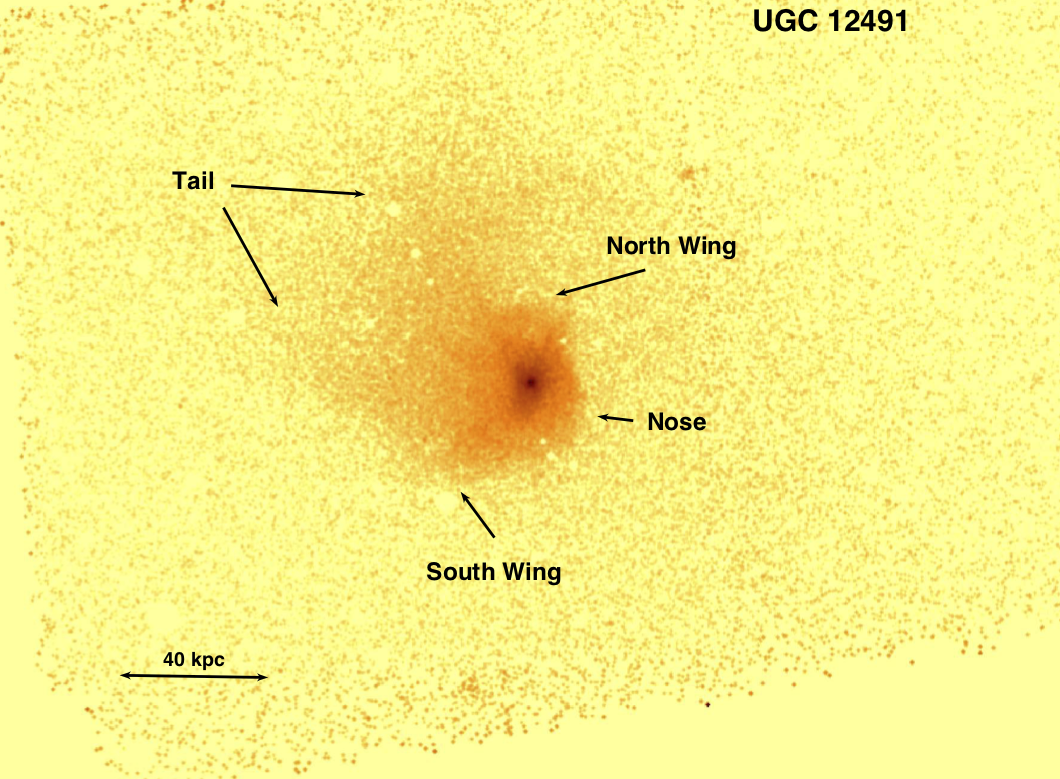}
\includegraphics[width=5in]{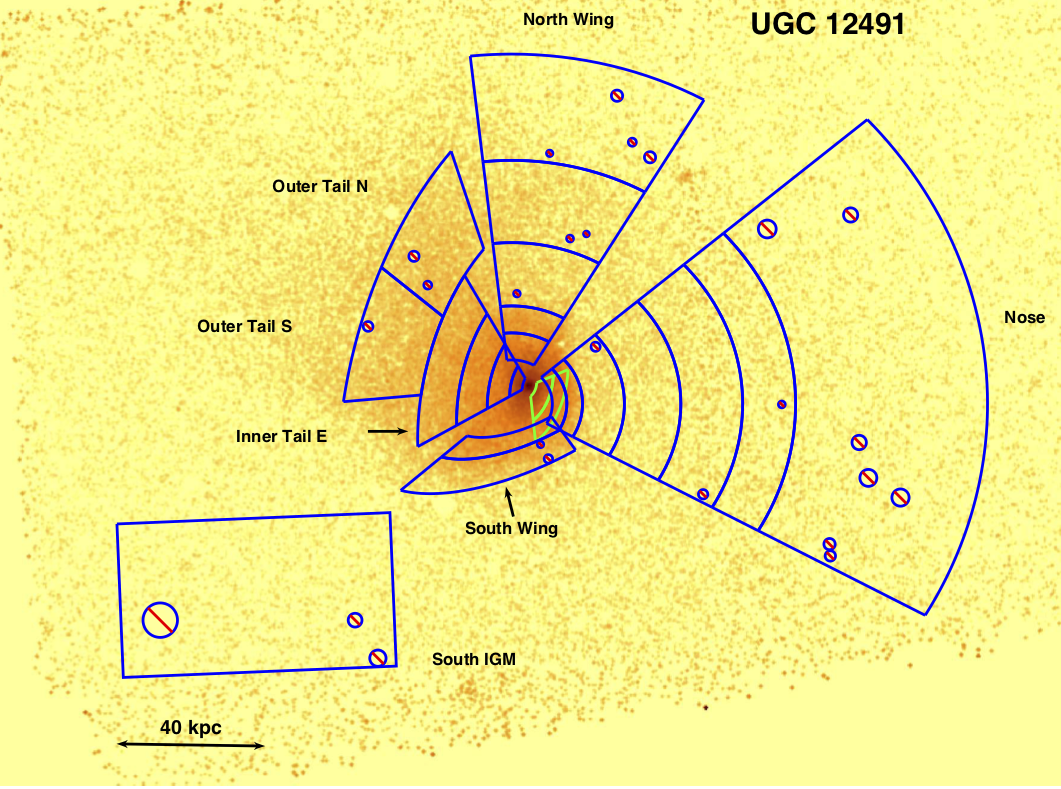}
\caption{(Upper panel) $0.5-2$\,keV background-subtracted, 
exposure-corrected co-added {\it Chandra} X-ray image of UGC\,12491
with a  combined useful exposure of $147$\,ks. Point sources other than
the central AGN  have been excluded and the image is 
smoothed by a  $1\as$ Gaussian kernel. 
$1\,{\rm  pixel} =0.984\as \times 0.984\as$. North is up, west is
to the right, and $40\,{\rm kpc} = 1.85$\,arcmin.  (Lower panel) Sectors for 
the X-ray surface brightness profile analysis from 
  Table \protect\ref{tab:ugc12491sectors} and the spectral regions defined in
  Appendix B 
are  overlaid on the {\em Chandra } image of UGC\,12491. The Inner W regions, 
chosen for comparison with the Inner Tail E regions, are shown in green.
The box spectral region (South IGM) is chosen in the direction of 
NGC\,7618 to measure the gas properties of the IGM between the merging 
groups. The resulting surface brightness profiles are shown in 
Fig. \protect\ref{fig:ugc12491profs}.
}
\label{fig:ugcanalysis}
\end{center}
\end{figure*}

\subsection{UGC\,12491}
\label{sec:ugcanalysisintro}

In the upper panel of Figure \ref{fig:ugcanalysis}, we present 
the high resolution $0.5-2$\,keV background-subtracted, 
exposure-corrected co-added {\it Chandra} X-ray image of
UGC\,12491 with a total effective exposure of $147$\,ks. The image 
has been binned by two instrument pixels such that 
$1\,{\rm bin} = 0.984\as \times 0.984\as$ and smoothed with 
a $1''$ Gaussian kernel.  The peak X-ray surface 
brightness is coincident with the optical center of the
galaxy. However, the observed diffuse X-ray 
gas distribution about the galaxy's center is highly asymmetric.
To the west there is a bright, elliptical leading edge that for 
simplicity we label as {\em Nose}. The center of the ellipse that traces 
the Nose is  displaced $6.8$\,kpc to the southeast 
of the optical center.  This displacement of the brighter (denser) 
galaxy gas to the south of the galaxy's center is consistent with 
merger-induced sloshing. The leading edge is not smooth, but appears
 ragged or scalloped on scales of $\sim 6 - 10\as$ ($\sim 2 - 3$\,kpc). 

An apparently  bifurcated X-ray tail emerges from UGC\,12491 toward 
the east and northeast. The southern segment of the tail extends 
$\sim 60$\,kpc to the east before 
either fading into the surrounding IGM, or joining the 
northern tail segment that, in projection,  forms a tightly wound spiral 
to the north and west around the galaxy.  
Excess emission,  measured  relative to that from the IGM to the southeast 
away from the tail, is  observed directly to the west (in front) of the Nose. 
This excess emission  appears
to connect smoothly to the westward arc of the tail and so may be a 
continuation of its westward spiral (see also Fig. \ref{fig:merge}).  

 There are two wing  features (Fig. \ref{fig:ugcanalysis} top panel).
The South Wing appears in projection arc-like with its wing {\em tip}
extending just south of the southern tail segment. The
North Wing is boxy, with two bright surface brightness edges. The innermost 
edge of the North Wing follows an extension of the ellipse that traces the 
Nose edge. The second bright edge may separate disturbed galaxy gas, 
possibly in the process of being stripped by KHI, from gas in 
the tightly wound tail, seen in projection.
In the lower panel of Figure \ref{fig:ugcanalysis}, we superpose 
regions on the X-ray image of UGC\,12491, which we use in the following 
subsections to construct and parameterize surface brightness profiles 
across edges and along part of the tail and to measure hot gas temperatures
and abundances in these regions of interest.

\begin{figure*}[htb!]
\begin{center}
\includegraphics[width=3in,angle=0]{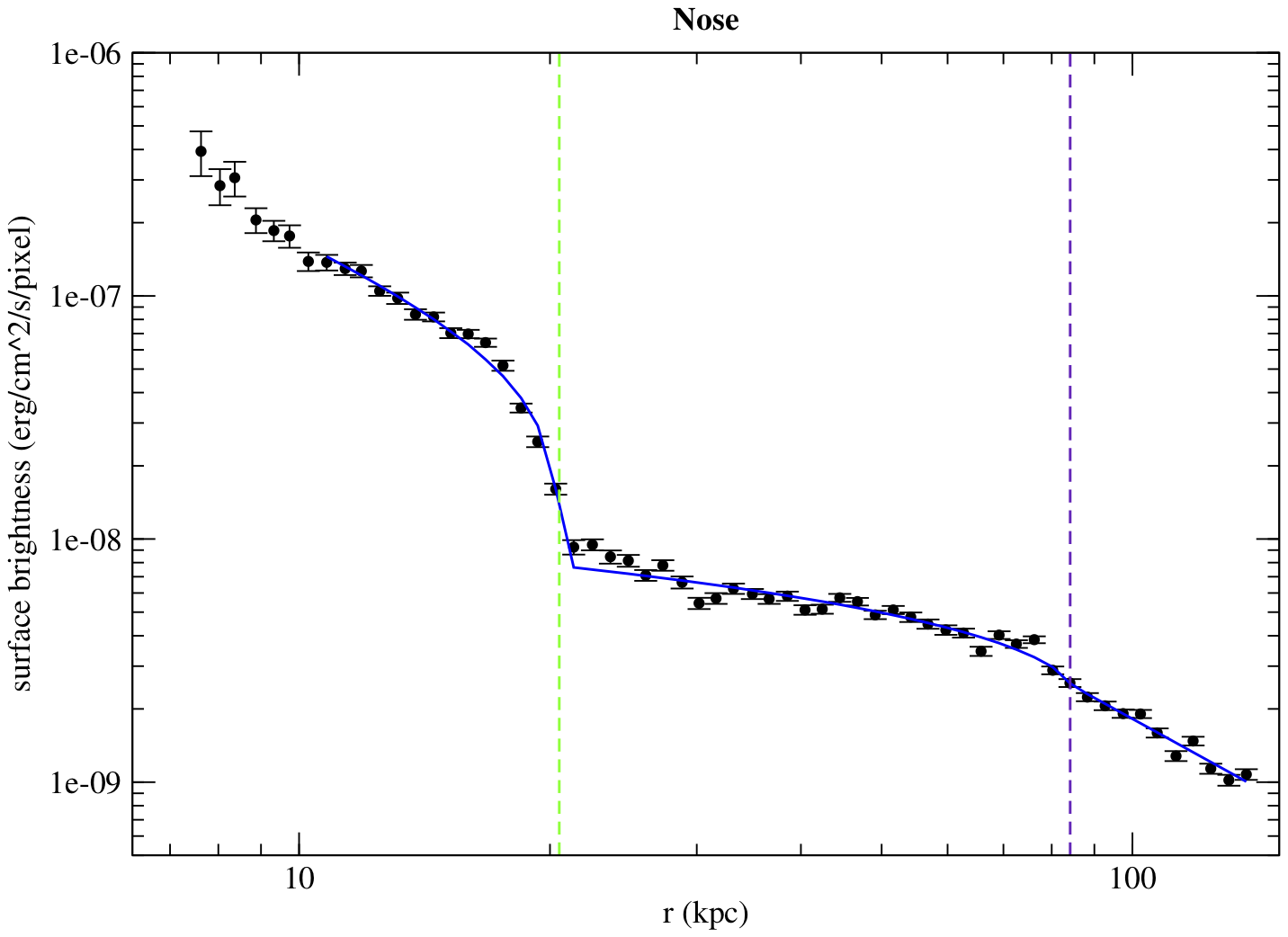}
\includegraphics[width=3in,angle=0]{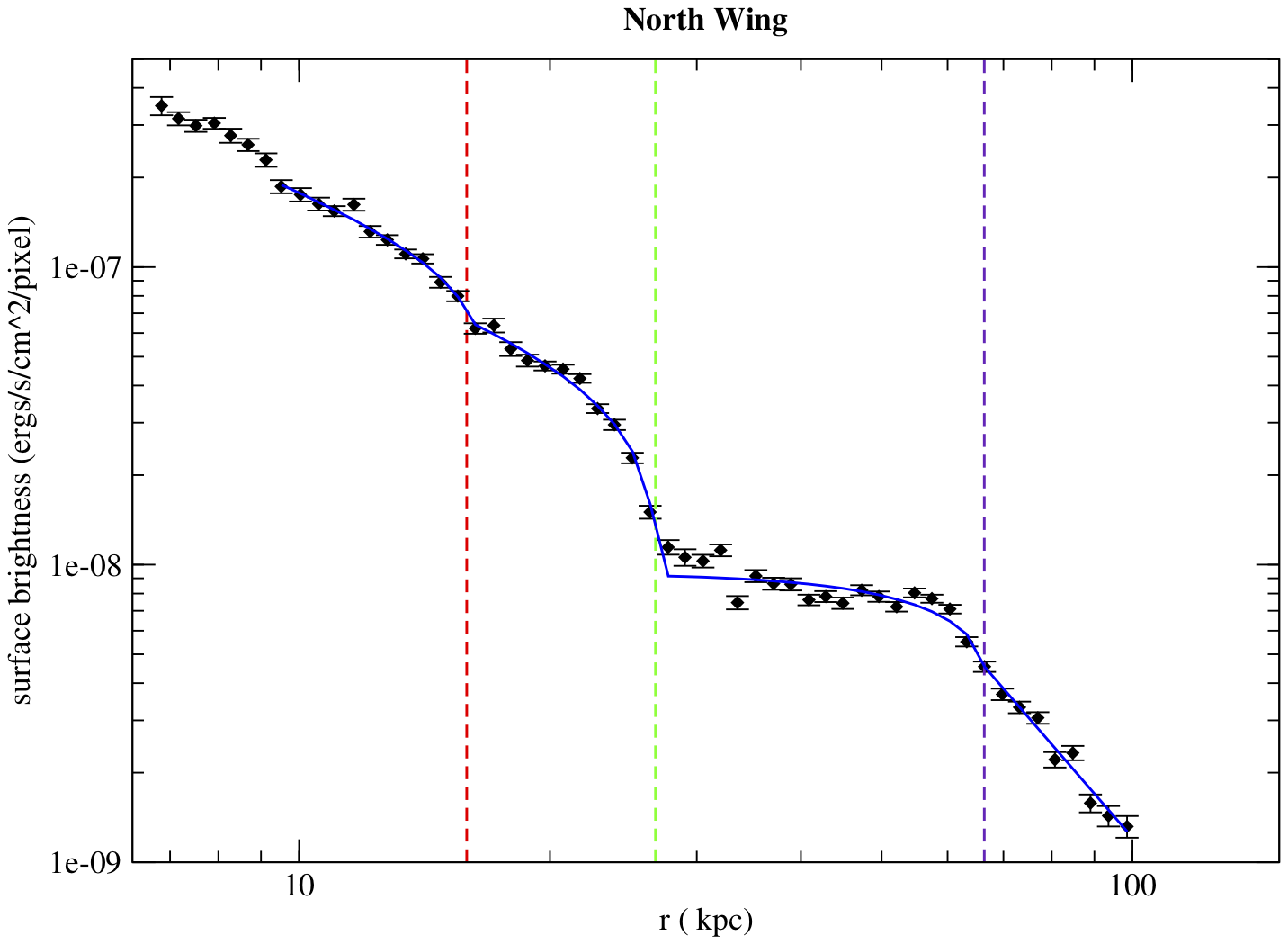}
\includegraphics[width=3in,angle=0]{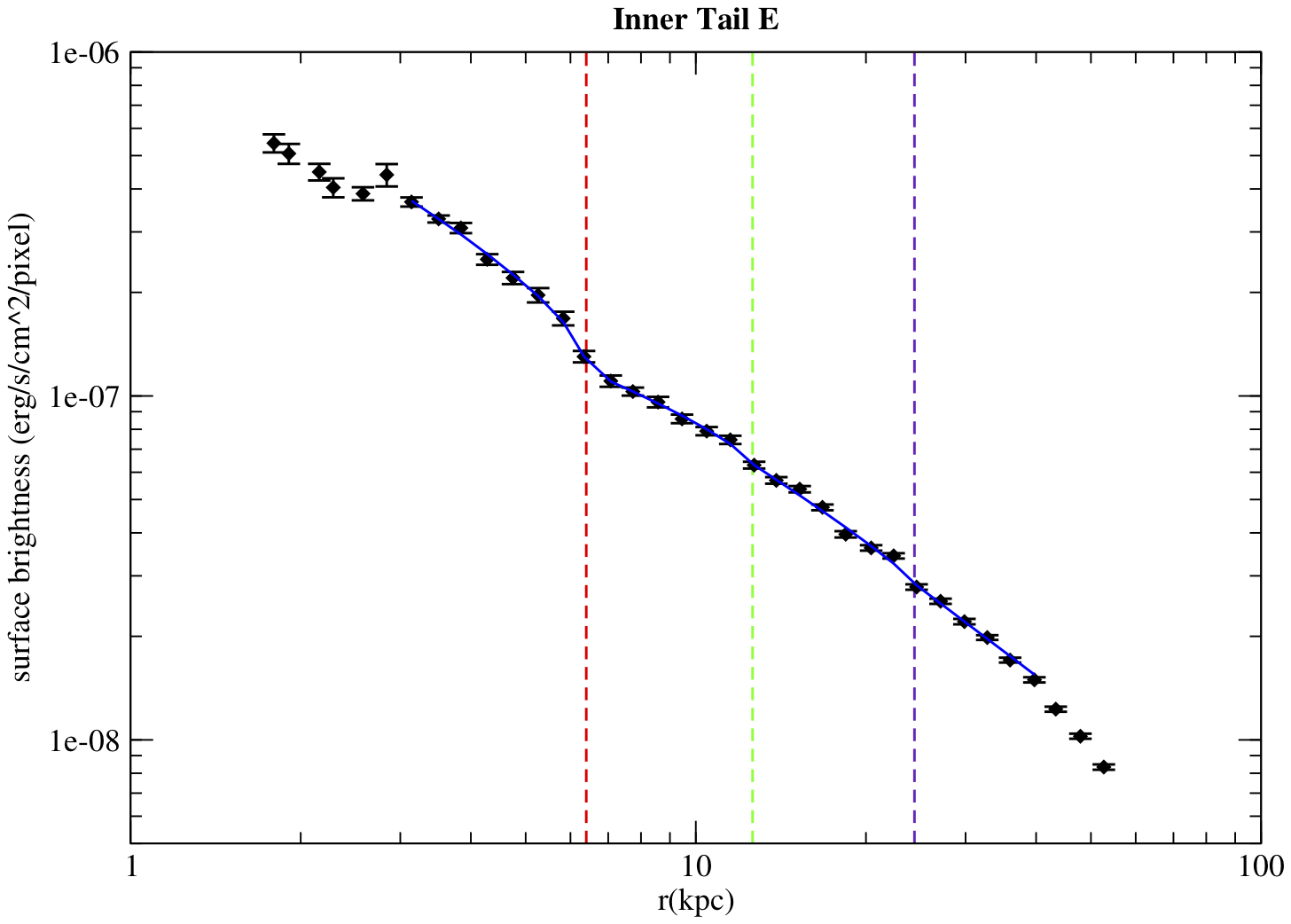}
\includegraphics[width=3in,angle=0]{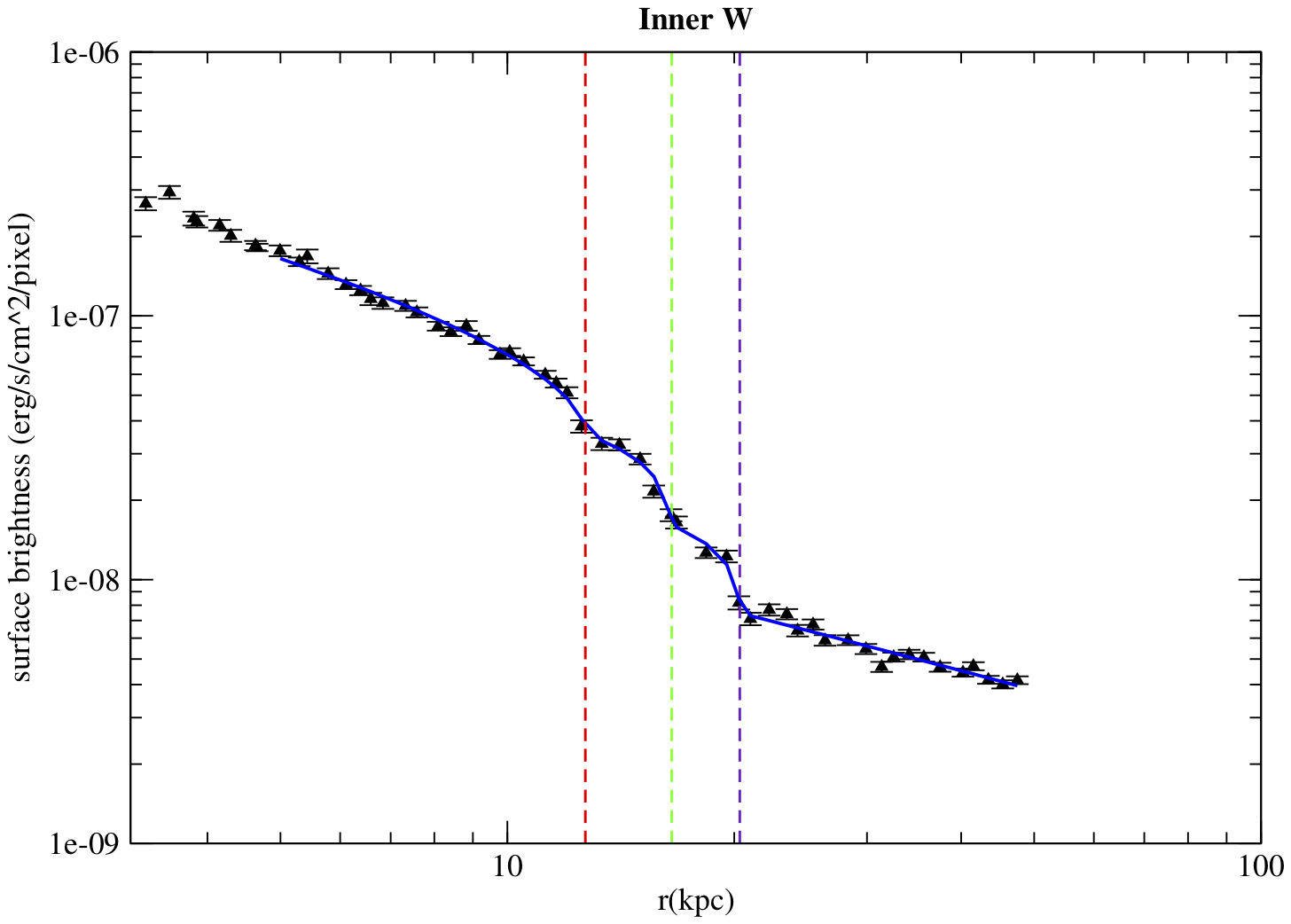}
\caption{ $0.5-2$\,keV X-ray surface brightness profiles of UGC\,12491
  constrained to sectors shown in the lower panel of 
 Fig. \protect\ref{fig:ugcanalysis} and listed in 
 Table \protect\ref{tab:ugc12491sectors}. Each profile is  superposed
 with the power law density model fits (solid line) listed in 
  Tables \protect\ref{tab:ugc12491alpha}  
  and \protect\ref{tab:ugc12491edges}. Vertical dashed lines denote
  the location of edges. From top left to bottom right the profile
  regions are the 
  Nose with edges at $20.52$\,kpc (green) and $84.2$\,kpc (magenta); 
  North Wing with edges at $15.9$\,kpc (red), $26.8$\,kpc (green),
  and $66.4$\,kpc (magenta); Inner Tail E with edges at $6.4$\,kpc (red),
  $12.6$\,kpc (green), and $24.4$\,kpc (magenta); and Inner W with edges at
  $12.7$\,kpc (red), $16.5$\,kpc (green), and $20.3$\,kpc (magenta).}
\label{fig:ugc12491profs}
\end{center}
\end{figure*}

\begin{deluxetable}{ccc}
\tablewidth{0pc}
\tablecaption{UGC\,12491 Profile Sectors}\label{tab:ugc12491sectors}
\tablehead{\colhead{Profile Label} & \colhead{A} &
  \colhead{B} \\
  & (degree) & (degree) }
\startdata
Nose   & $333$  & $38.2$ \\  
North  Wing   &$57.2$  & $96.75$  \\  
Inner Tail E  & $120.4$ & $208.5$  \\ 
Inner W   &$276$ & $22.4$ \\ 
\enddata
\tablecomments{Each sector is centered at the center of the bounding
 ellipse for the corresponding region. See \S\ref{sec:specdefs} 
of Appendix B for complete definitions of the bounding ellipses for these 
regions.
Each sector subtends the angle from angle A to angle B, with  all
angles measured counterclockwise from the west. 
}
\end{deluxetable}

\subsubsection{X-ray Surface Brightness Profiles}
\label{sec:ugc12491profanyl}

To characterize the complex gas morphologies observed in UGC\,12491, 
we construct X-ray surface brightness profiles from the background-subtracted, 
exposure-corrected, co-added  $0.5-2$\,keV image shown in 
Figure \ref{fig:ugcanalysis}. We identify sectors containing one
or more surface brightness discontinuities ({\em edges}) and define for each
sector of interest a bounding ellipse that traces the morphology of
a dominant edge-like feature in that sector (see the lower panel of
Fig. \ref{fig:ugcanalysis} and Table \ref{tab:ugc12491sectors}). 
The complete definition of each bounding ellipse used in our analysis of
UGC\,12491 is given in Appendix B. 
The surface brightness profile within each sector is 
then constructed by measuring the $0.5-2$\,keV  X-ray surface
brightness in elliptical annuli concentric to the bounding ellipse, 
using increasing logarithmic radial steps to move across the edges
from small to large radii.  

We use four sectors to study the gas features of interest in
UGC\,12491. We use a profile to the west across the sharp {\em Nose}
feature to probe the gas properties within and outside the leading edge, 
and  a profile predominantly to the north to study the hot gas properties 
across multiple edges in the boxy 
North Wing feature ({\em North Wing}). We use a profile to the east 
( {\em Inner Tail E} ) centered on UGC\,12491's X-ray peak to study the 
properties of hot gas as one moves from the central region of the core to the 
inner edge of the tail before the tail appears, in projection,  to bifurcate 
and curve to the north and west. For comparison,
 we use a similar sector to the west ({\em Inner W}) also centered on 
the galaxy's X-ray peak. The resulting surface brightness profiles, 
shown in Figure \ref{fig:ugc12491profs}, confirm the complex, multiple 
edge-like gas features visible in the images.  

\begin{deluxetable}{ccccc}
\tablewidth{0pc}
\tablecaption{UGC\,12491 Density Model Power-Law Indices}\label{tab:ugc12491alpha}
\tablehead{ \colhead{Profile Label} & \colhead{$\alpha_1$} &
  \colhead{$\alpha_2$} & \colhead{$\alpha_3$} & \colhead{$\alpha_4$}}
\startdata
 Nose &$-1.19^{+0.15}_{-0.09}$ &$-0.47^{+0.05}_{-0.06}$
 & $-1.44^{+0.09}_{-0.09}$ & \ldots \\
North Wing & $-0.86^{+0.22}_{-0.22}$ & $-0.83^{+0.15}_{-0.13} $ & 
   $0.10^{+0.13}_{-0.06}$ & $-2.12^{+0.12}_{-0.12}$  \\
 Inner Tail E &$-0.92^{+0.12}_{-0.15}$ &$-0.67^{+0.21}_{-0.06}$
      &$-0.97^{+0.15}_{-0.07}$   & $-1.12^{+0.01}_{-0.10}$  \\
Inner W$^a$  &$-0.83$ &$-0.14$ &$-0.055$ & $-0.84$ \\
\enddata
\tablecomments{Density model power law indices $\alpha$  labeled by
  increasing integers from inside the innermost edge to outside the
  outermost edge.  Superscript {\em a} indicates the
  uncertainty analysis failed to converge due to too few data
  points between the edges. All other quoted uncertainties are $90\%$
  CL.}
\end{deluxetable}
 
\begin{deluxetable*}{ccccccc}
\tablewidth{0pc}
\tablecaption{{UGC\,12491 Density Model Edges and Jumps}\label{tab:ugc12491edges}}
\tablehead{ \colhead{Profile Label} & \colhead{$r_{e1}$} & \colhead{$j_{e1}$}
 & \colhead{$r_{e2}$} & \colhead{$j_{e2}$} 
 & \colhead{$r_{e3}$} & \colhead{$j_{e3}$} \\
 & (kpc)&   & (kpc)  &   & (kpc) &  }
\startdata
Nose & $20.52^{+0.04}_{-0.10}$ & $4. 52^{+0.33}_{-0.28}$ & 
      $84.2^{+1.3}_{-4.5}$ & $1.16^{+0.04}_{-0.04}$ & \ldots &\ldots \\
North Wing & $15.9^{+0.04}_{-0.23}$ & $1.35^{+0.11}_{-0.11}$ & 
    $26.8^{+0.05}_{-0.20}$ & $3.82^{+0.11}_{-0.12}$ & 
    $66.4^{+0.7}_{-3.3}$ & $1.20^{+0.06}_{-0.07}$ \\
Inner Tail E$^a$ & $6.4^{+0.20}_{-0.10}$ & $1.49^{+0.16}_{-0.12}$ & 
     $12.6^{+1.4}_{-2.6}$ & $1.08^{+0.07}_{-0.06}$ & $24.4$ & 
     $1.06^{+0.06}_{-0.02}$ \\
Inner W$^a$ & $12.69$ & $1.4$ & $16.52$ & $ 1.63$ & $20.34$ & $2.11$ \\
\enddata
\tablecomments{Best fit radial position of edges $r_{ek}$ and 
jumps $j_{ek}$  labeled by increasing integers $k$ from the innermost 
edge to the outermost edge. Note that for low mass galaxy groups the cooling 
function is sensitive to the metal abundance in the gas, such that the 
jumps $j_{ek}$ listed here are only proportional to the 
effective gas density ratios. Superscript {\em a} indicates the
uncertainty analysis failed 
to converge for one or more variables. All quoted uncertainties 
are $90\%$ CL.}
\end{deluxetable*}

We fit the surface brightness profiles in Figure
\ref{fig:ugc12491profs} by
integrating simple power-law gas density models along the LOS
with a {\em jump} in the density at each surface brightness
discontinuity ({\em edge}),  using a  multivariate $\chi^2$ minimization
algorithm developed by  M. Markevitch (private communication). We fit 
across multiple edges iteratively, first fitting for the location 
and jump of the outermost edge allowing the edge location, jump and 
outer and inner power law indices to vary. 
We then move inward, fixing the location
of the outer edge and the power law index outside that edge, but allowing 
the outer edge jump, next inner edge location and its jump,
as well as the power law indices on each side of the next inner 
edge, to vary. We proceed similarly for additional edges, 
 as we move to smaller radii.
A detailed description of the density models and of the fitting procedure 
used to model the profiles with multiple edges in 
Figure \ref{fig:ugc12491profs} is given in Appendix A. Within and near the
galaxy, these spherically symmetric models may be a reasonable
approximation to the hot gas density. However, at large radii,  spherical 
symmetry may no longer hold, as one looks through tail gas that is 
either fully or partially  stripped.  
 These models should then be interpreted as descriptive, 
providing a simple parameterization that can be used to quantitatively 
compare the observational data with simulations.

The best density model fits to the surface
brightness profiles are overlaid on the data in 
Figure \ref{fig:ugc12491profs}. Table \ref{tab:ugc12491alpha} 
lists the best-fit density power-law indices. 
Table \ref{tab:ugc12491edges} gives the edge locations and 
jumps. All given  uncertainties are $90\%$\,CL.   For two profiles, 
Inner Tail E and Inner W, the  uncertainty 
in one or more variables is poorly determined. The shape of the
single-edge model does not provide a good description of the Inner
Tail E profile. 
While the multiple-edge model does describe the Inner Tail E surface 
brightness profile well, the locations of  the two outer edges in the 
multi-edge model are not well  determined because the fitted jumps for 
those edges are close to $1$. Thus, the shape of the profile across the two 
outer edges of the Inner Tail E region behaves more like a broken power 
law than the characteristic shape expected for a contact discontinuity.  
For the Inner W profile, the uncertainties in the model parameters are 
difficult to measure because 
there are too few radial bins ($ \leq 4$) between the edges for the error 
analysis to converge. However, the data do not support decreasing 
the profile step size further in an attempt to better define the profile 
between these edges because of the 
increase in scatter in the measured surface 
brightness when measured in these smaller bins.  

We note that the model jumps listed in 
Table~\ref{tab:ugc12491edges} are only proportional to the ratio 
of the hot gas density inside the edge to that outside the edge, 
with a proportionality constant
given by $\sqrt{\Lambda_o/\Lambda_i}$ where $\Lambda_o$
($\Lambda_i$)  is the X-ray cooling function for gas outside (inside)
the edge (see, for example,  Machacek \etal 2005, 2006). As discussed
by Vikhlinin \etal (2001), for massive X-ray clusters where the gas 
temperature is much greater than $2$\, keV, gas cooling is dominated by 
the continuum emission, such that the X-ray cooling functions are only
weakly dependent on gas temperatures and abundances. Thus, for massive 
clusters, the ratio of the cooling functions on either side of the edge 
is $\sim 1$, and the fitted edge jump in these models is equal to the ratio 
of the hot gas density across the edge. However, for low-mass galaxy
groups with gas temperatures $\sim 1$\,keV, as is the case for
the UGC\,12491 and NGC\,7618 galaxy groups, line cooling 
dominates and the  cooling functions are 
very sensitive to the metal abundances in the gas. Since the metal 
abundances can vary significantly between galaxy gas and 
the IGM, the ratio of cooling functions cannot be ignored.  We
thus refer to the density model fit parameters $j_{ek}$ in 
Table \ref{tab:ugc12491edges}
 simply as jumps and defer 
the determination of the corresponding gas density ratios until after 
the discussion of the spectral properties of the hot gas along the 
profiles and in other features of interest. 

\subsubsection{Spectral Analysis} 
\label{sec:ugcspectralfits}

We fit the X-ray spectra extracted from the regions shown in the 
lower panel of Figure \ref{fig:ugcanalysis} using an absorbed APEC
model (Smith \etal 2000) for thermal emission from  an optically thin, 
collisionally ionized plasma with fixed Galactic hydrogen column density
$nH = 1.18 \times 10^{21}$\cmc (Dickey \& Lockman 1990) and 
photoionization absorption crosssections taken from Verner \etal (1996).
We note that Kraft \etal (2006) determined from ACIS GIS
observations that the X-ray emission from hot gas in this merger
extends well beyond the optical disks of the dominant group galaxies,
and as also shown in Figure \ref{fig:merge}, the emission does not
follow the optical
  light. For such gas-rich low-mass galaxy groups dominated by
  emission $\sim 1- 2$\,keV, the contribution to the X-ray
  emission from stellar sources such as low mass x-ray binaries
  can be neglected (Revnivtsev \etal 2008).  

Our results are given in Table \ref{tab:ugc12491spec}. 
The temperatures of hot gas within UGC\,12491 (see the regions Nose1,
Nose2, North Wing 1, 2, and S Wing 1 regions, both of the Inner W
regions, and  Inner E Tail 1, 2) are $\sim 1$\,keV with abundances 
$\sim 0.3-0.5\,\Zs$. The temperature of group gas to the south, measured in 
region the South IGM region (see Table \ref{tab:ugc12491spec}) is higher 
($1.30^{+0.06}_{-0.09}$\,keV) and the 
 abundance lower ($0.18^{+0.10}_{-0.07}\,\Zs$),  
as one would expect for group gas gravitationally bound to its  more massive
dark matter halo before the group gas has become
enriched. 

The measured abundance is also low ($\sim 0.14^{+0.07}_{-0.60 }\,\Zs$) for the 
$1.27^{+0.07}_{-0.09}$\,keV gas  outside the outermost  edge of 
the North Wing in the   
North Wing 5 region (see Table \ref{tab:ugc12491spec}).
For the North Wing regions between the second and third edges 
(North Wing 3 , 4 regions
in Table \ref{tab:ugc12491spec}),
the hot gas temperature is also consistent with that of the group gas to 
the south. However, the abundances in the North Wing3 and North Wing4 regions
are modestly higher, albeit with large 
uncertainties, and more similar to those  measured in 
the Outer Tail N and S regions. The highest gas temperatures and most
dramatic change in temperature occur to the west, across the inner 
bright edge of the Nose region, where temperatures increase from 
$1.05^{+0.02}_{-0.03}$\,keV inside the edge to $1.5-1.6$\,keV 
outside the edge. 
This gas temperature is higher than elsewhere in the group IGM,
suggesting that the gas may have been heated either by subsonic  
compression or shocks at some point in the merger history. 
Gas temperatures along the tail to the east 
increase from $1.0$\,keV  at a radial distance $r \sim 3$\,kpc  from the 
galaxy center, consistent with hot gas within the galaxy,  to $1.2$\,keV 
at $r\sim 23$\,kpc, consistent within uncertainties with the temperature 
of the group IGM measured in the South IGM and North Wing 5 regions 
(see Table \ref{tab:ugc12491spec}). 
Although the uncertainties are large, the metal
abundances in the tail are higher than those measured away from the
tail in the group IGM. This is consistent with the tail being composed of
higher abundance stripped or partially stripped galaxy gas viewed
through and/or being mixed with the lower abundance group IGM. 
The  measured Fe abundances for UGC\,12491 are in agreement,  within 
uncertainties, with Suzaku measurements 
(Mitsuishi \etal 2014). They find abundances of $0.2-0.4\,\Zs$ in the 
central regions of UGC\,12491, declining to $\sim 0.1-0.2\,\Zs$ to the 
southwest, midway between UGC\,12491 and NGC\,7618.

\begin{deluxetable*}{ccccccc}
\tablewidth{0pc}
\tablecaption{UGC\,12491 Spectral Models}\label{tab:ugc12491spec}
\tablehead{\colhead{Region} & \colhead{$kT$} &\colhead{A} &\colhead{$N_{\rm{APEC}}$}
  &\colhead{Flux} &\colhead{$\Lambda$} & \colhead{$\chi^2{\rm dof}^{-1}$} \\
 & (keV) &($\Zs$) &($10^{-4}$\,cm$^{-5}$) & ($10^{-13}$\ergscm) &
 ($10^{-23}$\,ergs\,cm$^3$\,s$^{-1}$) & }
\startdata
Nose1         &$1.02^{+0.02}_{-0.03}$ &$0.50^{+0.18}_{-0.11}$ &$0.7841$  &$1.0265$ &$1.35$  &$70.8/68$ \\
Nose2         &$1.05^{+0.02}_{-0.03}$ &$0.30^{+0.10}_{-0.07}$ &$0.8331$  &$0.74289$ &$0.92$ & $68.3/54$ \\
Nose3        &$1.52^{+0.13}_{-0.16}$ &$0.13^{+0.09}_{-0.06}$ &$1.1707$  &$0.6185$  &$0.55$ &$101.3/68$ \\
Nose4         &$1.58^{+0.12}_{-0.15}$ &$0.17^{+0.10}_{-0.07}$ &$1.3542$  &$0.74826$ &$0.57$ &$80.3/86$  \\
Nose5         &$1.56^{+0.11}_{-0.14}$ &$0.20^{+0.10}_{-0.08}$ &$1.4863$  &$0.85885$ &$0.60$ &$85.6/84$  \\ 
Nose6         &$1.46^{+0.14}_{-0.12}$ &$0.19^{+0.13}_{-0.08}$ &$1.3986$  &$0.81482$  &$0.60$ &$77.3/68$  \\         
Nose7         &$1.49^{+0.11}_{-0.13}$ &$0.14^{+0.07}_{-0.05}$ &$4.2165$  &$2.2628$  &$0.56$ &$227.8/227$ \\ 
North Wing1   &$0.97^{+0.03}_{-0.02}$ &$0.37^{+0.11}_{-0.08}$ &$1.1925$  &$1.2978$  &$1.13$ & $83.3/79$ \\   
North Wing2   &$1.04^{+0.02}_{-0.03}$ &$0.27^{+0.08}_{-0.06}$ &$0.9713$  &$0.81426$ &$0.87$  &$67.2/58$  \\
North Wing3   &$1.30^{+0.05}_{-0.05}$ &$0.21^{+0.07}_{-0.06}$ &$1.2446$  &$0.79074$  &$0.66$ & $74.1/72$  \\
North Wing4   &$1.25^{+0.03}_{-0.05}$ &$0.33^{+0.10}_{-0.08}$ &$1.4146$  &$1.1401$  &$0.83$ &$115.7/113$ \\
North Wing5   &$1.27^{+0.07}_{-0.09}$ &$0.14^{+0.07}_{-0.06}$ &$1.4805$  &$0.82193$  &$0.57$ &$93.3/110$ \\ 
Inner W1      &$1.02^{+0.02}_{-0.02}$ &$0.37^{+0.09}_{-0.07}$  &$1.3729$ &$1.4412$  &$1.09$  &$73.2/85$ \\
Inner W2      &$1.03^{+0.02}_{-0.02}$ &$0.40^{+0.11}_{-0.09}$  &$1.0118$ &$1.1009$  &$1.13$  &$81.4/74$ \\
Inner Tail E1 &$0.93^{+0.03}_{-0.03}$ &$0.26^{+0.08}_{-0.06}$ &$1.2754$  &$1.1382$ &$0.92$  &$75.3/66$   \\
Inner Tail E2 &$1.04^{+0.02}_{-0.02}$ &$0.34^{+0.08}_{-0.06}$ &$1.5448$  &$1.5017$  &$1.01$ &$112.7/94$  \\
Inner Tail E3 &$1.12^{+0.03}_{-0.04}$ &$0.33^{+0.08}_{-0.07}$ &$2.1723$  &$1.9398$  &$0.92$ &$124.7/117$ \\
Inner Tail E4 &$1.21^{+0.03}_{-0.02}$ &$0.55^{+0.12}_{-0.10}$ &$1.9291$  &$2.1802$ &$1.17$  &$124.8/138$ \\
Outer Tail N &$1.16^{+0.04}_{-0.05}$ &$0.31^{+0.08}_{-0.07}$ &$1.9460$  &$1.626$&$0.87$  &$151.2/122$ \\ 
Outer Tail S &$1.18^{+0.04}_{-0.04}$ &$0.27^{+0.06}_{-0.05}$ &$2.3682$  &$1.8089$&$0.79$ & $171.2/129$  \\
South Wing1   &$1.00^{+0.02}_{-0.02}$ &$0.38^{+0.11}_{-0.08}$ &$1.2303$ &$1.3336$ &$1.12$  & $77.4/84$  \\
South Wing2   &$1.11^{+0.05}_{-0.04}$ &$0.24^{+0.09}_{-0.05}$ &$1.5667$  &$1.1929$  &$0.79$ &$86.4/85$   \\
South IGM     &$1.30^{+0.06}_{-0.09}$ &$0.18^{+0.10}_{-0.07}$ &$1.4386$ &$0.87145$  &$0.63$ & $205/175$   \\
\enddata\tablecomments{Spectra were modeled using the absorbed APEC model
  (phabs $\times$ apec) with Galactic absorption fixed at 
$1.18 \times  10^{21}$\cmc (Dickey \& Lockman 1990). 
  All regions except the South IGM region are elliptical panda regions,
  shown in the
lower panel of Fig. \protect\ref{fig:ugcanalysis} and defined in 
Appendix B. Region numbers increase from small to large radii. 
Columns from left
to right: ($1$) region name, ($2$) temperature, ($3$) abundance, ($4$) 
unabsorbed X-ray flux in the $0.5-2$\,keV energy band, ($5$) APEC model
normalization, ($6$) measured cooling function $\Lambda$ in the
$0.5-2$\,keV energy band (see Eq. \protect\ref{eq:lambda}), and ($7$)
$\chi^2{\rm dof}^{-1}$ for the model fit.}
\end{deluxetable*}

\subsubsection{Edge Analysis: Effective Gas Densities and Pressures}
\label{sec:ugcedgeanyl}

We use the surface brightness profile fits combined with the spectral
models for the gas to determine gas density and pressure ratios across the 
surface brightness edges to gain insight into the physical processes
responsible for the disturbed gas morphology observed in this merger. 
The jumps $j_{ek}$ and edge locations $r_{ek}$ are labeled by increasing 
integers $k$ from the innermost to the outermost edge and  
listed in Table \ref{tab:ugc12491edges}. The jumps $j_{ek}$ are  
related to the ratio $n_i/n_o$ of the electron gas densities $n_i$ inside 
the edge ($r<r_{ek}$) and $n_o$ outside the edge ($r> r_{ek}$)
 through the ratio of the corresponding $0.5-2$\,keV X-ray cooling 
functions $\Lambda_i$ ($\Lambda_o$) for hot gas inside (outside) the 
edge, respectively.
\begin{equation}
 j_{ek}^2 = \frac{\Lambda_i}{\Lambda_o}(\frac{n_i}{n_o})^2
\label{eq:jump}
\end{equation}
The cooling function $\Lambda$ is inferred from the APEC model
normalization and spectral fit properties
of the gas in the regions on either side of the edge using: 
\begin{equation}
 \Lambda = \frac{10^{-14}FD_L^2}{N_{\rm{APEC}}[D_A(1+z)]^2} 
\label{eq:lambda}
\end{equation}
where $F$ is the unabsorbed model flux in the $0.5-2$\,keV energy 
band, required to match that of the surface brightness profile, 
$N_{\rm{APEC}}$ is the
APEC model normalization, $z$ is the redshift, $D_L(D_A)$ are the
luminosity (angular size) distances, respectively, and 
$D_L^2 \sim [D_A(1+z)]^2$ for $z << 1$. Values for 
$N_{\rm{APEC}}$, $F$, and $\Lambda$ for each spectral region  
are also given in  Table \ref{tab:ugc12491spec}. 

In Table \ref{tab:ugc12491ratios}, 
we identify each edge by the profile label and edge location and 
calculate the corresponding electron density $n_i/n_o$, gas 
temperature $T_i/T_o$, and gas pressure $p_i/p_o = (n_iT_i)/(n_oT_o)$ 
ratios across each edge. For each edge in 
Table \ref{tab:ugc12491ratios} 
we list the two spectral regions on either side of the edge  
shown in the lower panel of Figure \ref{fig:ugcanalysis} with the region 
number within the sector increasing with increasing radius. 
For all but three edges, the effective thermal
pressure ratio measured across the edge is $1$ within
uncertainties.  This is expected for edges produced by gas
sloshing. The modest overpressure ($p_i/p_o= 1.40^{+0.23}_{-0.18}$ )
for gas at the edge $6.4$\,kpc east of the galaxy center may be a
sign of ongoing stripping of galaxy gas into the tail. 

The effective thermal pressure ratios of $2.7^{+0.2}_{-0.3}$ and 
$2.4^{+0.5}_{-0.4}$ across the boxy North Wing edge at $r=26.8$\,kpc and 
the Nose at $r=20.5$\,kpc, respectively, are more difficult to explain. 
In each case, the sharp decrease in surface 
brightness coincides with a transition from galaxy gas to a region of 
excess X-ray emission that, in projection, may be composed of remnant tail 
and group gas. The density distribution of the more dense remnant tail 
may not be well described by a spherically symmetric model  because of 
the arc-shaped morphology induced as the remnant tail 
outruns its parent galaxy and is carried out sidewise by the
conservation of angular momentum during the galaxy's approach to apogee. 
A quantitative understanding of what the spherically symmetric  effective 
fits imply about the physical three-dimensional density distribution and 
composition of gas in these regions requires that the same analysis be 
applied to mock Chandra X-ray images produced from simulation data matched 
to this merger. However, this is beyond the scope of the current paper.

\begin{deluxetable}{ccccc}
\tablewidth{0pc}
\tablecaption{UGC\,12491 Edge Analysis}\label{tab:ugc12491ratios}
\tablehead{ \colhead{Profile} & \colhead{$r_e$} &
  \colhead{$n_i/n_o$} & \colhead{$T_i/T_o$} & \colhead{$p_i/p_o$} \\
   & (kpc) & & &  }
\startdata
Nose (2,3)         &$20.52$ & $3.49^{+0.26}_{-0.21}$ & $0.69^{+0.10}_{-0.07}$ & $2.41^{+0.54}_{-0.38}$ \\
Nose (6,7)         &$84.2$  &$1.20^{+0.04}_{-0.04}$  & $0.98^{+0.11}_{-0.14}$ & $1.18^{+0.30}_{-0.21}$ \\
North Wing (1,2)   & $15.9$ & $1.19^{+0.09}_{-0.10}$ &$0.93^{+0.06}_{-0.03}$ & $1.11^{+0.16}_{-0.13}$ \\
North Wing (2,3)   & $26.8$ &$3.33^{+0.09}_{-0.11}$  &$0.80^{+0.05}_{-0.05}$ & $2.66^{+0.24}_{-0.25}$ \\
North Wing (4,5)   & $66.4$ & $1.00^{+0.04}_{-0.06}$ &$0.98^{+0.11}_{-0.08}$  & $0.98^{+0.15}_{-0.14}$   \\
Inner Tail E (1,2) & $6.4$  & $1.56^{+0.17}_{-0.08}$ &$0.89^{+0.05}_{-0.04} $ & $1.40^{+0.23}_{-0.18}$  \\ 
Inner Tail E (2,3) & $12.6$ & $1.03^{+0.06}_{-0.06}$ &$0.93^{+0.05}_{-0.04}$ & $0.95^{+0.12}_{-0.09}$ \\
Inner Tail E (3,4) & $24.4$ & $1.20^{+0.06}_{-0.03}$ &$0.93^{+0.04}_{-0.06}$ & $1.11^{+0.11}_{-0.09}$  \\
\enddata
\tablecomments{The numbers in parenthesis (i,o) following the region name 
indicate the spectral region inside ( $r < r_e$) and outside ($r > r_e$) 
 the corresponding edge located at radius $r_e$, respectively. Uncertainites 
for derived values assume  extremes in the $90\%$ CL uncertainties for 
measured properties.}  
\end{deluxetable}

\begin{figure*}
\begin{center}
\includegraphics[width=5in]{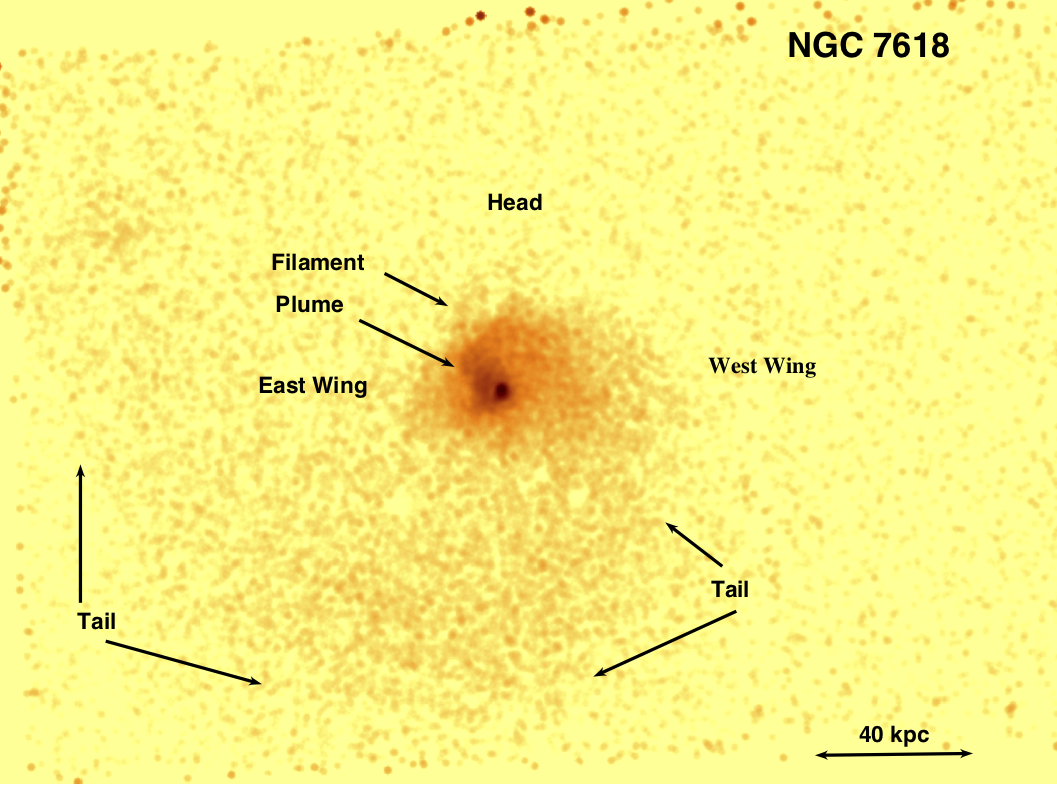}
\includegraphics[width=3in]{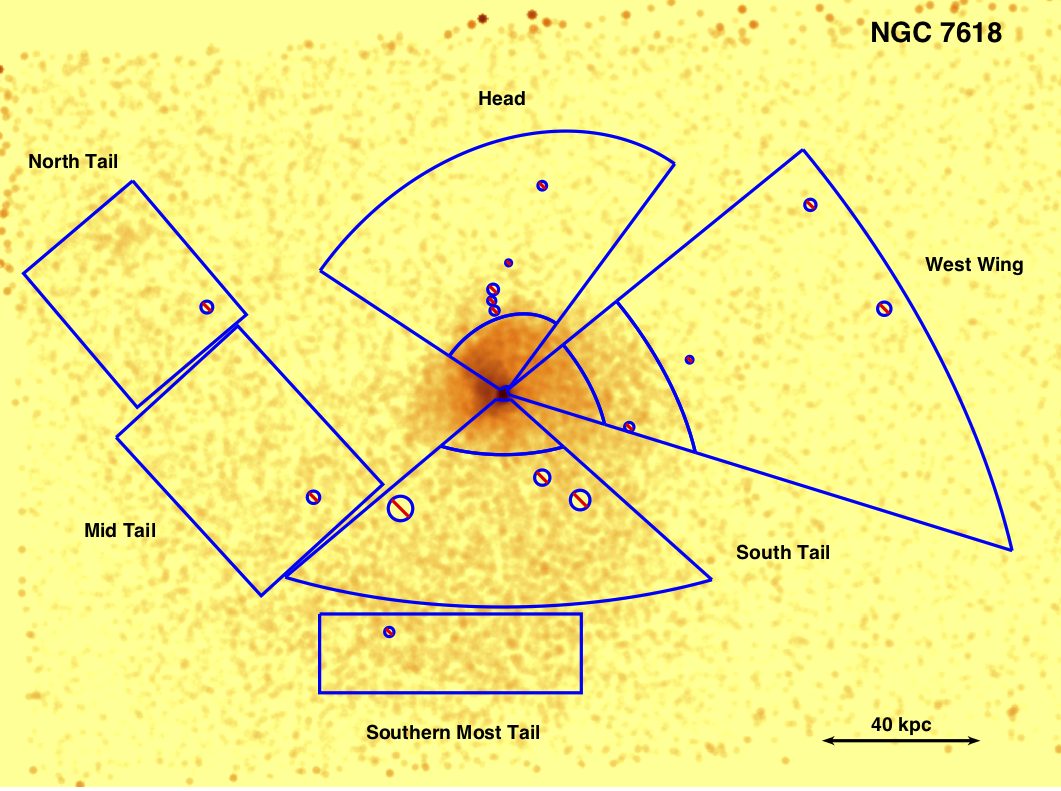}
\includegraphics[width=3in]{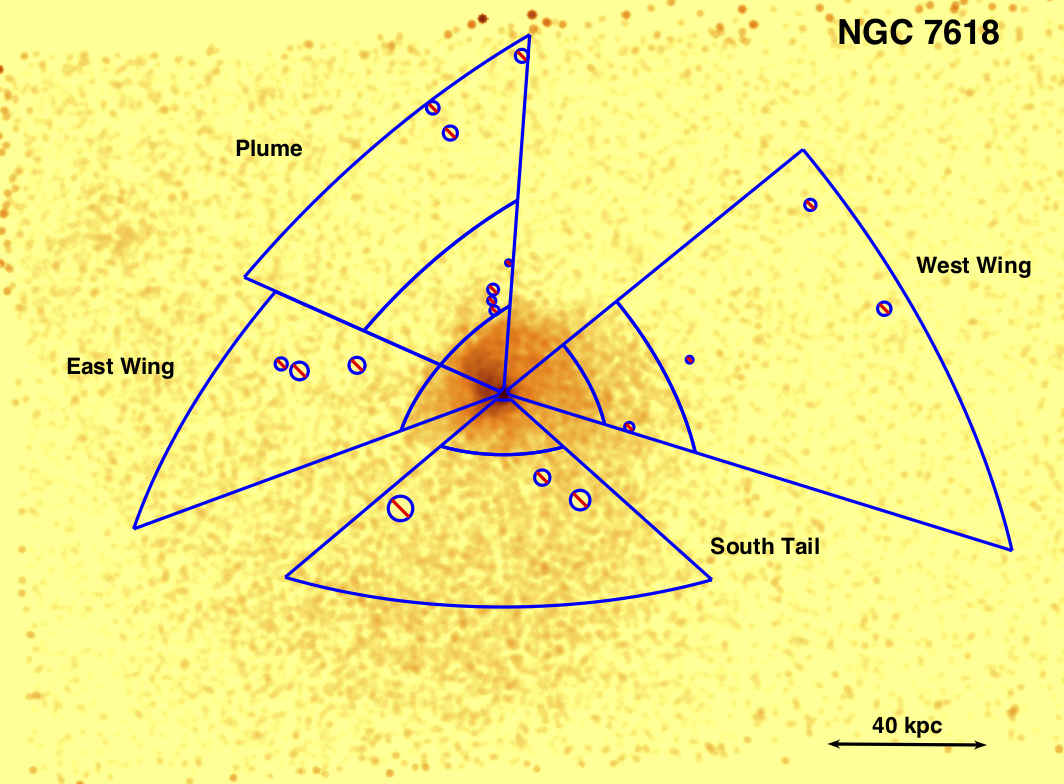}
\caption{({\it Upper panel}) $0.5-2$\,keV background-subtracted, 
exposure-corrected co-added {\it Chandra} X-ray image of NGC\,7618 
with features labeled. Point sources have been excluded and the image
has been smoothed with a $3\as$ Gaussian kernel to highlight faint
emission.  $1\,{\rm  pixel} =0\as.984 \times 0\as.984$ and the scale 
$40\,{\rm kpc}=1.85$\,arcmin. 
({\it Lower-left}) Sectors for the X-ray surface brightness profile
analysis and spectral regions emphasizing the head and tail features from 
\protect\S\ref{sec:n7618spectra} 
are overlaid on the {\it Chandra } image of NGC\,7618 
from the upper panel. ({\it Lower-right panel}) Sectors for the X-ray
surface brightness profile analysis 
and spectral regions emphasizing the northern plume and east wing features
are overlaid on the {\it Chandra} image from the upper panel.
 }
\label{fig:ngcanalysis}
\end{center}
\end{figure*}

\subsection{NGC\,7618}
\label{sec:ngcanalysisintro}

In the top panel of Figure \ref{fig:ngcanalysis}, we show the $0.5-2$\,kev 
background-subtracted, exposure-corrected, mosaicked {\it Chandra}
image of the NGC\,7618 galaxy subgroup. Point sources, other than the
nuclear point source, have been excluded. The image has $1\as$ pixels
and has been smoothed with a $3\as$ Gaussian kernel to highlight the
faint gas features. Asymmetric structure is observed in the hot 
gas in and surrounding NGC\,7618, and in spite of its lower 
($50.48$\,ks) total useful exposure, the gas features in NGC\,7618 are 
strikingly similar to those observed in UGC\,12491 
(shown in Fig. \ref{fig:ugcanalysis}). NGC\,7618 has a nose-like, 
leading surface brightness edge to the northeast (labeled Head in 
Fig. \ref{fig:ngcanalysis}), {\em wings} of emission  to the east and
west sides of the leading edge, and a long ($\sim 170$\,kpc) arced tail.   
At radii inside the leading edge, Figure \ref{fig:ngcanalysis} shows 
dense gas pushed to the northeast in a plume, before curving sharply
to the north and again to the west, creating a spiral with angular
(boxy) edges. This structure has been seen in previous images of NGC\,7618 
using {\it  Chandra} ObsId 7895 (Roediger \etal 2012a, Goulding
\etal 2016). The leading edge (Head) also is not smooth. A faint, narrow
($\sim 4$\,kpc wide) gas filament extends $11$\,kpc ($\sim 32\as$)
to the north, outside of the plume's boxy, spiral extension. 
 Moving radially out from the nucleus through the West Wing region in 
Figure \ref{fig:ngcanalysis}, we see two surface brightness edges (see also 
the profile analysis in Figure \ref{fig:n7618profs}).  The inner edge 
appears more closely associated with the Head region and inner spiral,
while the outer edge could either be stripping, due to hydrodynamic
instabilities, or gas from the tail seen in projection to curve around the
galaxy, first  to the south and then to the north. 

As in Figure \ref{fig:ugcanalysis} for UGC\,12491, the lower panels of Figure
\ref{fig:ngcanalysis} for NGC\,7618 outline the regions 
that are used to study  the X-ray
surface brightness profiles, X-ray spectra, and effective densities
and pressures across the edges in \S\ref{sec:ngcprofiles}, 
\S\ref{sec:n7618spectra}, and \S\ref{sec:ngcedges}, respectively, 
superposed on the image of NGC\,7618 shown in the upper panel.  
In the lower-left panel, we treat the Head region as a whole, using a sector
 with an opening angle of $93^\circ$ to 
average over the brighter plume, and we trace the northward segment
of the tail with rectangular regions. In the lower-right panel, we
isolate the plume using a narrower sector with an opening angle of $58^\circ$,
 and we also choose a sector to construct the surface brightness profile 
across the edge of the East Wing. Note, however,  that the  outer
spectral region of the East Wing sector overlaps part of the Mid-Tail region.
The West Wing and South Tail regions are the same in each figure.

\begin{deluxetable}{ccc}
\tablewidth{0pc}
\tablecaption{NGC\,7618 Profile Sectors}\label{tab:n7618sectors}
\tablehead{\colhead{Profile Label} & \colhead{A} &
  \colhead{B} \\
  & (degree) & (degree) }
\startdata
Head         &$53.4$   &$146.4$  \\
Plume        & $85.7$  & $155.8$ \\
West Wing    & $343.7$   &$399$   \\
East Wing    & $155.9$ & $200$ \\
South Tail   & $220$  & $318$  \\
\enddata
\tablecomments{All sectors are centered at the X-ray/optical peak
 $\alpha=23^{h}19^{m}47^{s}.22,\delta=+42^{\circ}51^{'}09.5\as$. 
Each sector subtends the angle measured counterclockwise from angle A 
to angle B. All angles are measured counterclockwise from the west. 
}
\end{deluxetable}

\subsubsection{X-ray Surface Brightness Profiles}
\label{sec:ngcprofiles}

To characterize the complex gas morphologies in NGC\,7618, we construct 
X-ray surface brightness profiles from the $0.5-2$\,keV 
background-subtracted, exposure-corrected, co-added {\it Chandra} 
image binned by $2 \times 2$ instrument pixels, such that 
$1 {\rm bin} =0.\as984 \times 0.\as984$
(similar to that  shown in Fig.  \ref{fig:ngcanalysis}, but without
Gaussian smoothing), using logarithmic
radial steps constrained to lie in the sectors defined 
in Table \ref{tab:n7618sectors} 
and shown in the lower panels of Figure \ref{fig:ngcanalysis}. 

\begin{figure*}[htb!]
\begin{center}
\includegraphics[width=3in,angle=0]{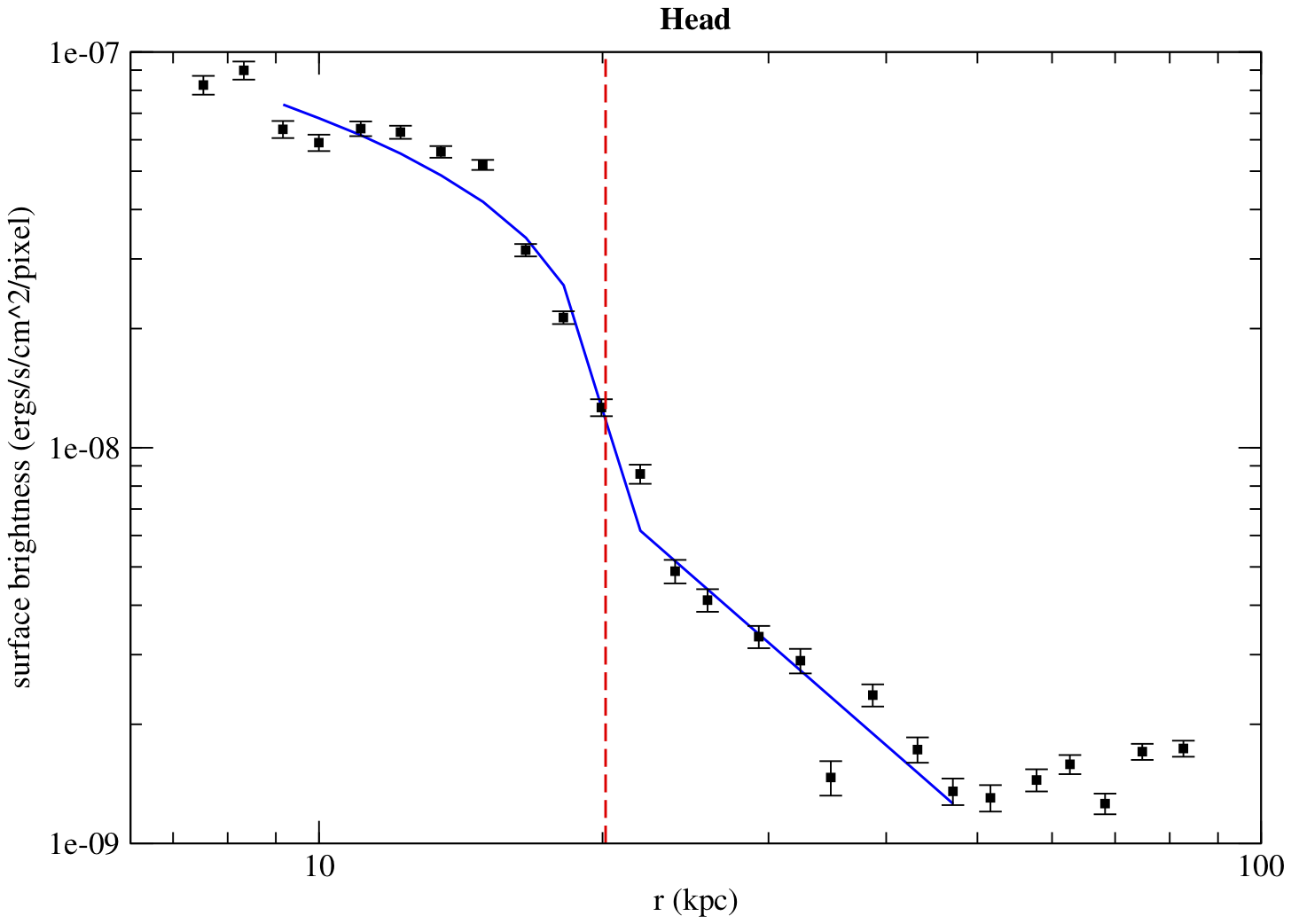}
\includegraphics[width=3in,angle=0]{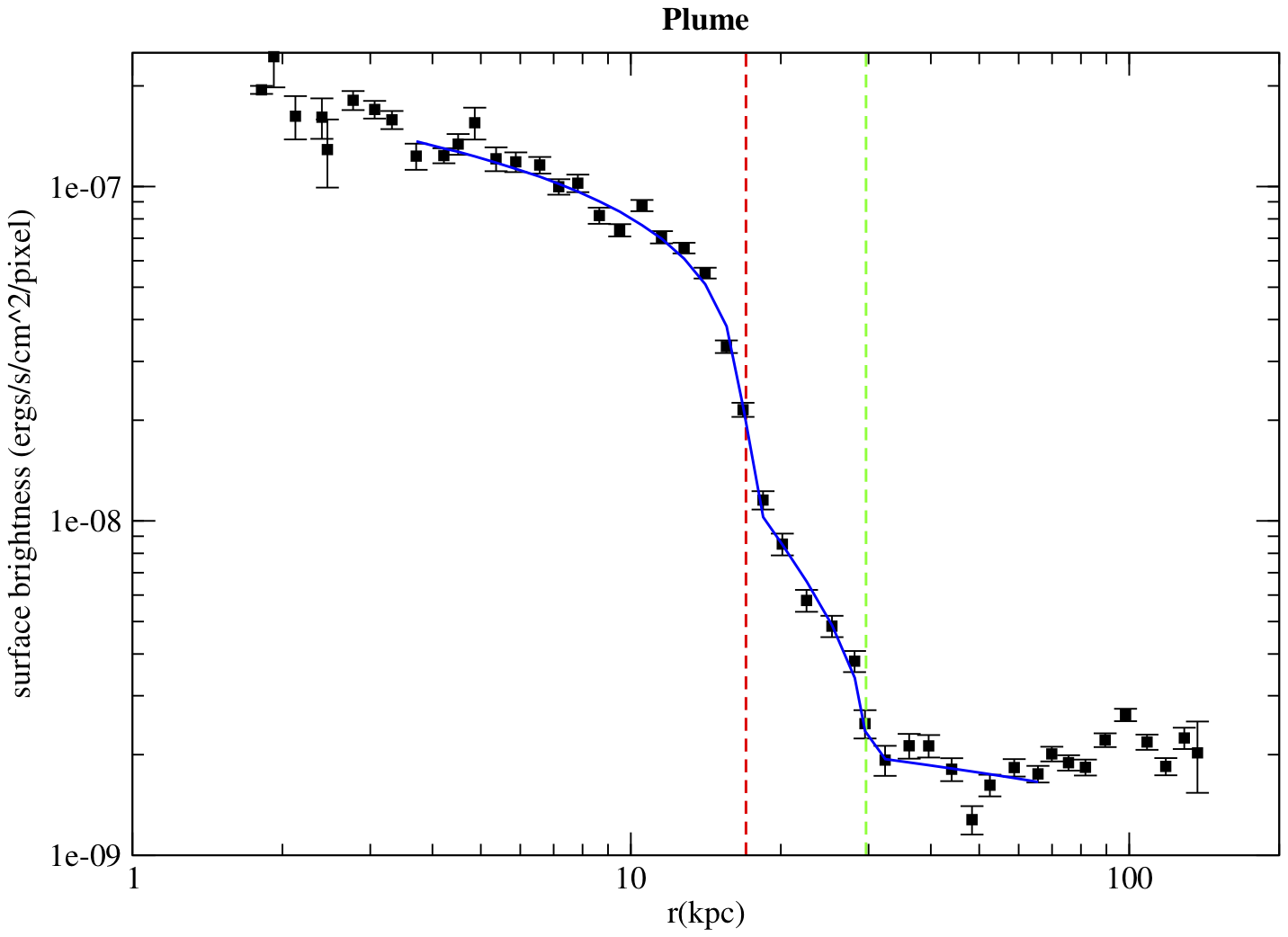}
\includegraphics[width=3in,angle=0]{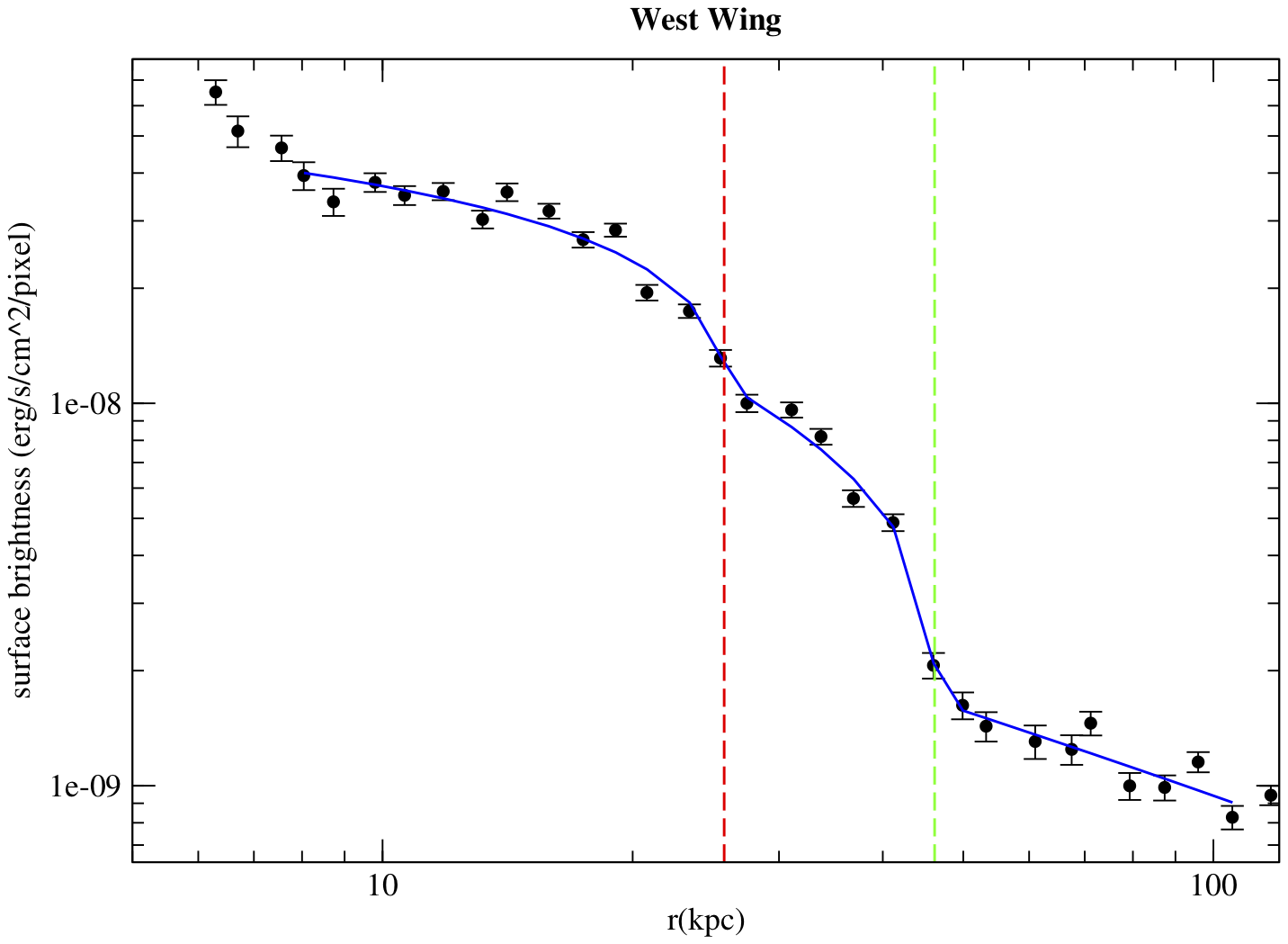}
\includegraphics[width=3in,angle=0]{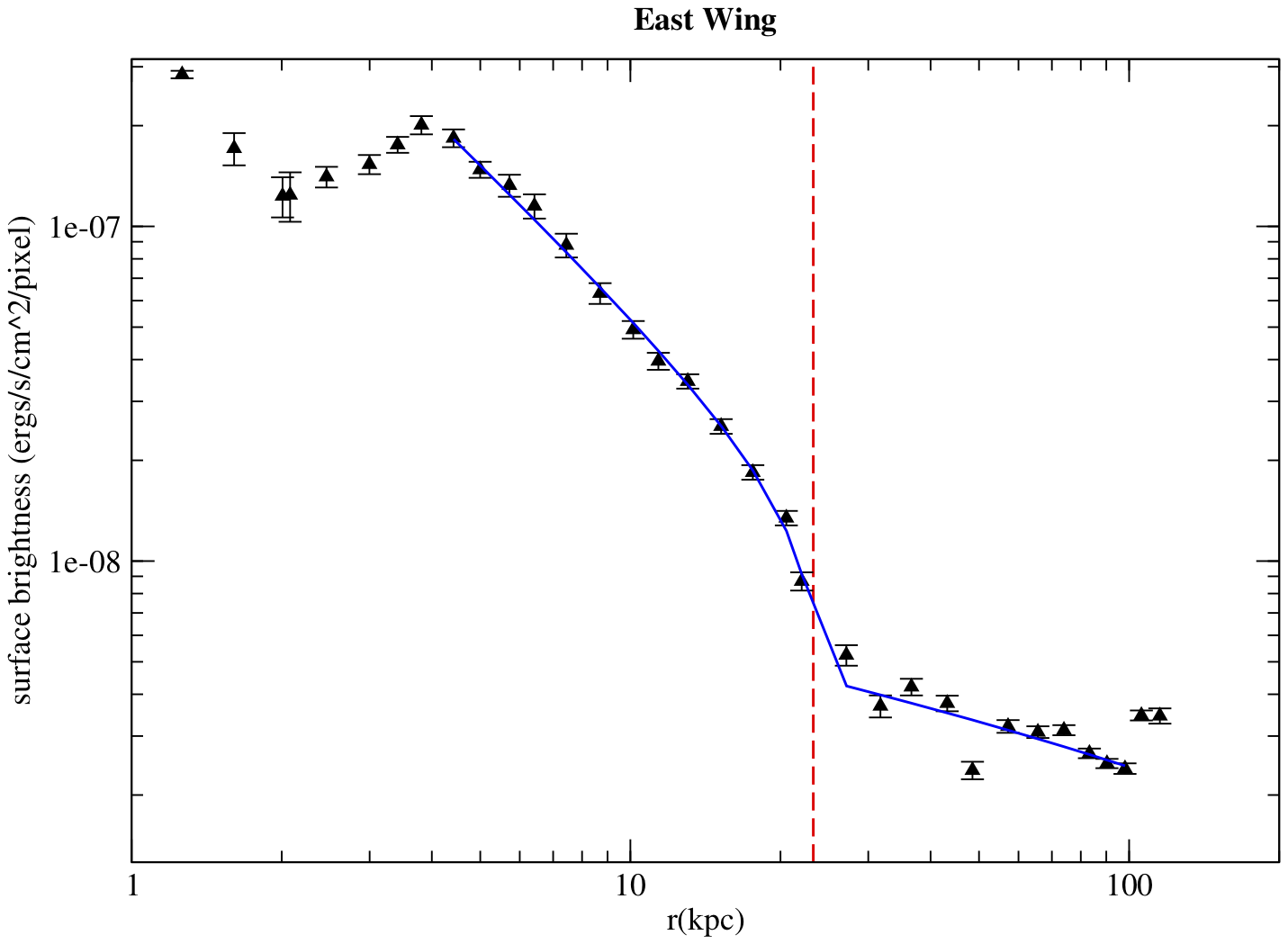}
\includegraphics[width=3in,angle=0]{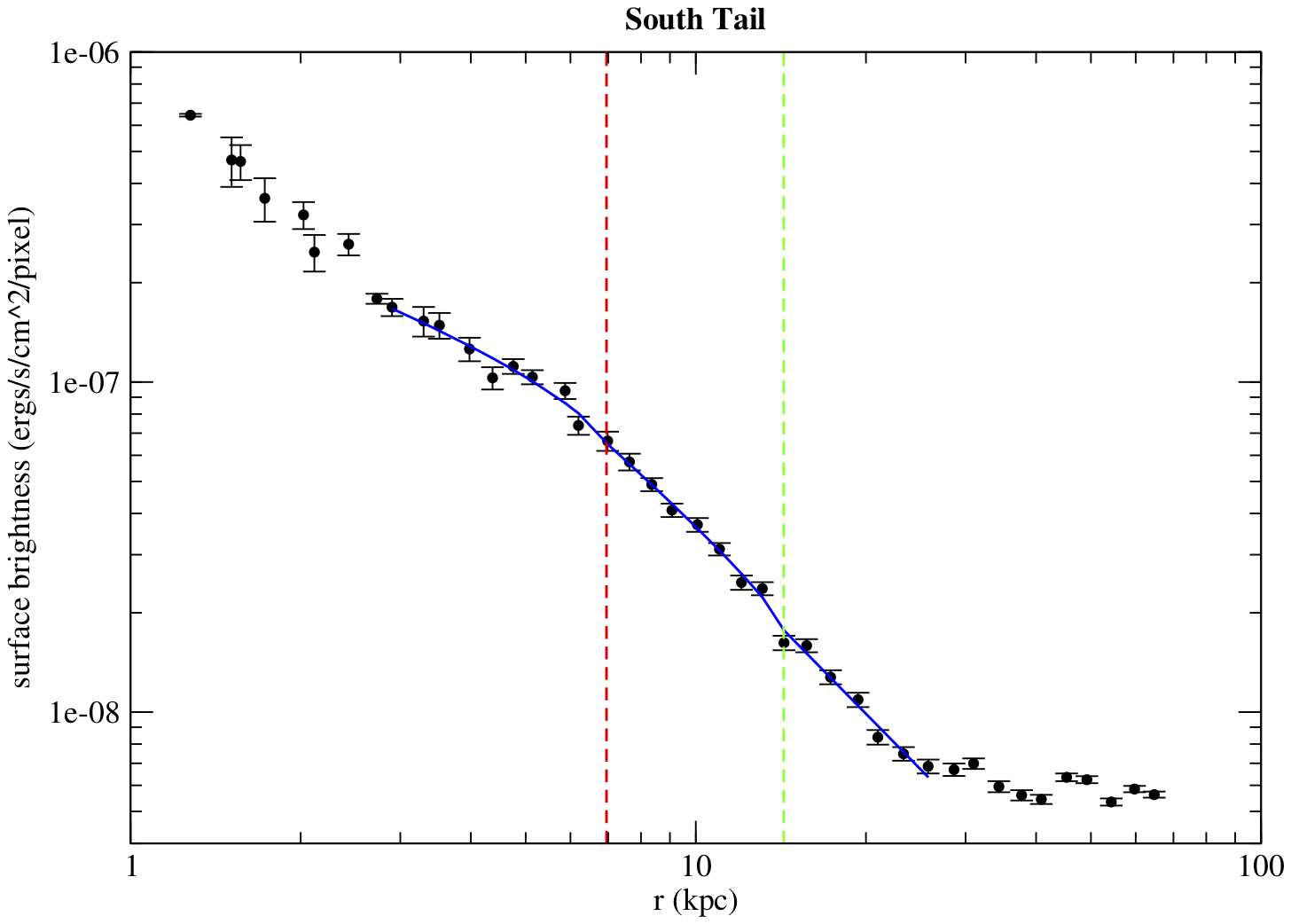}
\caption{ $0.5-2$\,keV X-ray surface brightness profiles of NGC\,7618
 labeled by and  constrained to the  sectors shown in the lower two panels of 
 Fig. \protect\ref{fig:ngcanalysis} and listed in 
 Table \protect\ref{tab:n7618sectors}.  
Each profile is  superposed  with the corresponding power-law density 
model fit (solid line) listed in Tables \protect\ref{tab:n7618alpha} 
and \protect\ref{tab:n7618edges}. 
Vertical dashed lines denote the location of edges. From upper left  
to lower right,  the profile regions are the  
  Head with the edge at $20.15$\,kpc (red); Plume with edges at $17.02$\,kpc
  (red) and $29.65$\,kpc (green); West Wing with edges at $25.77$\,kpc
  (red) and $46.18$\,kpc (green); East Wing with the
  edge at $23.27$\,kpc (red); and  
   South Tail with possible weak edges at $6.95$\,kpc (red) and $14.3$\,kpc
   (green).} 
\label{fig:n7618profs}
\end{center}
\end{figure*}

Following the same procedure as in 
\S\ref{sec:ugc12491profanyl} and described in  
Appendix A, we fit the surface brightness profiles
with a series of spherically symmetric power-law density models and
jumps at each observed surface brightness discontinuity. Due to the 
significantly lower effective exposure for NGC\,7618 caused by the 
solar flare contamination, fine structures within these profiles are  more
difficult to identify than in UGC\,12491 and the fits are less robust. 
These X-ray surface brightness profiles, overlaid with the density
model fits, are shown in Figure \ref{fig:n7618profs}. The best-fit
density model power law indices 
are given in Table \ref{tab:n7618alpha}, 
and the edge locations and jumps are given in Table \ref{tab:n7618edges}. 

\begin{deluxetable}{cccc}
\tablewidth{0pc}
\tablecaption{NGC\,7618 Density Model Power-Law Indices}\label{tab:n7618alpha}
\tablehead{ \colhead{Profile Label} & \colhead{$\alpha_1$} &
  \colhead{$\alpha_2$} & \colhead{$\alpha_3$} }
\startdata
 Head       &$-0.60^{+0.12}_{-0.08}$ &$-1.54^{+0.09}_{-0.12}$ &\ldots  \\
 Plume      &$-0.41^{+0.07}_{-0.05}$ &$-1.44^{+0.24}_{-0.55}$ & $-0.46^{+0.23}_{-0.05}$ \\
 West Wing  & $-0.34^{+0.10}_{-0.09}$ &$-0.83^{+0.21}_{-0.55}$ &$-0.82^{+0.09}_{-0.06}$  \\ 
 East Wing  &$-1.20^{+0.05}_{-0.04}$ & $-0.63^{+0.05}_{-0.04}$ & \ldots  \\ 
 South Tail     & $-0.72$ &$-1.22$ & $-1.37$  \\ 
\enddata
\tablecomments{Density model power-law indices $\alpha$ for the
  profiles shown in Fig. \ref{fig:n7618profs}. 
The power-law indices are labeled by
  increasing integers from inside the innermost edge to outside the
  outermost edge for each profile.  All quoted uncertainties 
 are $90\%$ CL. }
\end{deluxetable}
 
Both the West Wing and Plume profiles are best fit with  multiple 
edges. However, the Head profile, an average over the plume and galaxy
gas to the northwest of the plume inside the edge, supports only a
single edge model. The East Wing  is also best fit by a single edge model.   
 For the South Tail, the two edge model, shown
in Figure \ref{fig:n7618profs}, describes the data well, but the jumps
are weak ($\sim 1$) such that the profile is characterized by a
steepening of the power law index from $-0.72$ for radii $2<r<7$\,kpc to
$-1.22$ for radii  $7<r<14$\,kpc to $-1.37$ for radii $14<r<30$\,kpc.
As also seen for the Inner E Tail of UGC\,12491 in 
Figure \ref{fig:ugc12491profs}, neither a simple power law nor a  single 
edge density model provides a good description of the South Tail profile, 
and the uncertainties in the multiple edge model are poorly determined. 
We again defer the discussion of the effective gas density and
pressures until after we have mapped the hot gas temperatures and
abundances in and around NGC\,7618.

\subsubsection{Spectral Analysis}
\label{sec:n7618spectra}

As in \S\ref{sec:ugcspectralfits} we use an absorbed APEC model with
Galactic absorption to fit the hot gas temperature and heavy element
abundances in each region shown in the lower panels of 
Figure \ref{fig:ngcanalysis}.   Our results are given in 
Table \ref{tab:n7618spec}
with region number increasing with increasing radius for each sector.

\begin{deluxetable}{ccccc}
\tablewidth{0pc}
\tablecaption{NGC\,7618 Density Model Edges and Jumps}\label{tab:n7618edges}
\tablehead{ \colhead{Profile Label} & \colhead{$r_{e1}$}&\colhead{$j_{e1}$}&\colhead{$r_{e2}$}&\colhead{$j_{e2}$} \\
 & (kpc) &  & (kpc) &  }
\startdata
Head       &$20.15^{+0.01}_{-0.11}$ &$2.42^{+0.40}_{-0.68}$ &$\ldots$ &$\ldots$  \\
Plume      &$17.02^{+0.05}_{-0.13}$ &$2.37^{+0.28}_{-0.20}$ &$29.65^{+0.55}_{-0.15}$ & $3.27^{+0.35}_{-0.28}$\\
West Wing  &$25.8^{+0.3}_{-2.3}$ &$1.55^{+0.16}_{-0.29}$ &$46.18^{+0.25}_{-0.15}$ & $2.78^{+0.16}_{-0.20}$ \\ 
East Wing  &$23.3^{+0.07}_{-0.05}$ & $3.16^{+0.39}_{-0.29}$&$\ldots$ & \\ 
South  Tail &$6.95$ &$1.08$ &$14.3$ & $1.11$ \\ 
\enddata
\tablecomments{ Best-fit radial position of edges $r_{ek}$ and jumps 
$j_{ek}$ for $k={1,2}$ where $1,2$ label the inner and outer edges,
 respectively, for the profiles shown in Fig. \protect\ref{fig:n7618profs}.
All quoted uncertainties are $90\%$ CL.}  
\end{deluxetable} 

Spectral models for the Head1, East Wing1, West Wing1, and Plume1
regions find a hot gas temperature of $0.9-1$\,keV within the NGC\,7618
galaxy, consistent with previous measurements (Goulding \etal 2016). 
The best fit  heavy element abundances for hot galaxy gas are
$0.34^{+0.08}_{-0.06}\,\Zs$, $0.35^{+0.15}_{-0.10}\,\Zs$,
$0.30^{+0.11}_{-0.08}\,\Zs$, and $0.45^{+0.17}_{-0.11}\,\Zs$ for the  
Head1, East Wing1, West Wing1, and Plume1 regions, respectively.
These abundances 
 are consistent with each other within their $90\%$ CL uncertainties and 
are in agreement with Suzaku measurements by Mitsuishi \etal (2014). 
The central value for the metal abundance in the best-fit spectral model for 
the hot  gas in the bright plume (the Plume1 region) is
higher than in the other regions. Although the uncertainties are large, 
this may  suggest that the plume is 
composed of higher abundance gas dredged upward from the central
region of NGC\,7618 either by the merger-induced tidal interactions with
UGC\,12491 or by sloshing.

The West Wing3 and Plume3 regions are dominated by IGM gas. 
Their best fit temperatures
($1.32^{+0.14}_{-0.12}$\,keV and $1.35^{+0.16}_{-0.11}$\,keV) and
abundances ($0.11^{+0.07}_{-0.06}\,\Zs$ and $0.17^{+0.13}_{-0.08}\,\Zs$),
respectively, are consistent with the hot gas temperature and abundance
measured  for the IGM south of UGC\,12491 in the direction of
NGC\,7618 (see the South IGM region in Table \ref{tab:ugc12491spec}).
The Head2 region contains both the IGM gas and galaxy gas in the filament
and  possibly KHI-induced, irregular {\em rolls} along the edge. The
same is true of the Plume2 region. This accounts for temperatures that are 
intermediate between those of the IGM (see the West Wing3 and Plume3 regions) 
and galaxy gas (e.g. see the Plume 1, Head1, and West Wing1 regions).  This 
is also true for the East Wing2 region, which covers both IGM and tail gas. 

The most interesting temperature structure is found in the tail. 
The gas temperature decreases with increasing radius to the south from 
$1.07^{+0.04}_{-0.02}$\,keV in the bright innermost region to 
$0.75^{+0.03}_{-0.03}$\,keV in the southernmost region, before the tail
is observed to turn smoothly to the north and east. Measured
abundances in the same regions remain, within uncertainties, 
constant at $\sim 0.2 -0.4\,\Zs$, typical of that measured for other 
regions within NGC\,7618.  Thus, both the cool gas temperatures and 
the heavy element abundances support the hypothesis that the 
observed {\em tail} is composed of galaxy gas that has been displaced, 
but not necessarily gravitationally unbound, from the galaxy.
The temperature gradient to the south may be the result of tail gas cooling 
as it expands. However, it is intriguing to note that the spectral fit for the 
temperature and abundance for the  West Wing2 region is identical to that
in the South Tail2 region (see Table \ref{tab:n7618spec}). 
Since these regions are at similar mean radii from the center of NGC\,7618, 
this may suggest that West Wing2 is part of the same structure as 
South Tail2, forming a spiral feature that wraps continuously from the 
Head region into the tail, rather than an extension of West Wing1. In 
this latter 
interpretation of the tail morphology, the South Tail 
regions are oriented across the width of the tail rather than along its 
length, such that the measured temperatures reflect a temperature gradient 
across the width of the tail with higher temperature gas found  on the inner 
edge of the spiral feature closest to the galaxy. 
Finally, as the tail turns to the north and east (the Mid-Tail and
North Tail regions in Figure \ref{fig:ngcanalysis} and
Table \ref{tab:n7618spec}), 
the gas temperature slowly increases and the metal abundance
decreases toward the IGM values. This is likely the result of tail gas 
mixing with and/or viewed through proportionately more IGM gas.

\begin{deluxetable*}{ccccccc}
\tablewidth{0pc}
\tablecaption{NGC\,7618 Spectral Model Fits}\label{tab:n7618spec}
\tablehead{\colhead{Region} & \colhead{$kT$} &\colhead{A} & \colhead{Norm}
  &\colhead{Flux} &\colhead{$\Lambda$} & \colhead{$\chi^2/({\rm dof})$}\\
 & (keV) & ($\Zs$) & ($10^{-4}$\,cm$^{-5}$) & ($10^{-13}$\ergscm) &
 ($10^{-23}$\,ergs\,cm$^3$\,s$^{-1}$) & }
\startdata
Head1      & $0.92^{+0.02}_{-0.02}$  &$0.34^{+0.08}_{-0.06}$  &$2.91993$ &$3.1566$
&$1.12$ & $141/94$ \\
Head2      & $1.18^{+0.08}_{-0.12}$  &$0.11^{+0.06}_{-0.04}$
&$2.54809$ &$1.3649$ &$0.55$ &$143.8/126$ \\
Plume1$^a$     &$0.95^{+0.02}_{-0.03}$  &$0.45^{+0.17}_{-0.11}$ & $2.079$  &$2.660$
&$1.32$ & $50/54$ \\
Plume2     &$1.06^{+0.12}_{-0.07}$  &$0.12^{+0.08}_{-0.05}$   & $0.99045$  &$0.5622$
&$0.59$ &$45.4/41$ \\
Plume3     &$1.35^{+0.16}_{-0.11}$  &$0.17^{+0.13}_{-0.08}$  &$1.3863$   &$0.80223$   &$0.60$ & $104/88$\\
West Wing1 & $0.94^{+0.03}_{-0.03}$ & $0.30^{+0.11}_{-0.08}$  &$1.58476$ &$1.5405$   &$1.01$ & $72.3/59$\\
West Wing2 &$0.85^{+0.05}_{-0.04}$  & $0.21^{+0.10}_{-0.06}$  &$1.28025$   &$1.046$  &$0.85$ & $36.3/51$\\
West Wing3 &$1.32^{+0.14}_{-0.12}$ & $0.11^{+0.07}_{-0.06}$  &$2.24625$   &$1.1752$ &$0.54$ & $112.3/111$\\
East Wing1 & $0.98^{+0.03}_{-0.03}$&$0.35^{+0.15}_{-0.10}$  &$1.4279$ &$1.502$   &$1.09$   & $45.6/55$\\
East Wing2 & $1.15^{+0.07}_{-0.08}$&$0.22^{+0.10}_{-0.07}$ &$1.9604$&$1.3908$   &$0.73$   & $112.3/111$\\
South Tail1&$1.07^{+0.04}_{-0.02}$ &$0.29^{+0.08}_{-0.06}$  &$2.03174$ &$1.7401$   &$0.89$   & $80.2/70$\\
South Tail2&$0.85^{+0.02}_{-0.02}$ &$0.22^{+0.05}_{-0.04}$&$4.30845$ &$3.6458$   &$0.88$   &$160.5/154$ \\
SMost Tail &$0.75^{+0.03}_{-0.03}$ &$0.25^{+0.13}_{-0.07}$  &$1.77932$   &$1.6352$   &$0.95$   & $87.2/75$\\
Mid-Tail   &$1.01^{+0.03}_{-0.04}$ & $0.17^{+0.06}_{-0.04}$ &$2.4649$ &$1.6566$  &$0.70$   & $132/112$\\
North Tail &$1.06^{+0.08}_{-0.05}$ &$0.15^{+0.06}_{-0.04}$  & $1.77672$  &$1.0875$  &$0.63$   & $86.7/78$\\
\enddata
\tablecomments{Spectra were modeled using an absorbed APEC model with
Galactic absorption fixed. Superscript {\em a} based on ObsID 7895 only.}
\end{deluxetable*}

\subsubsection{Edge Analysis: Effective Gas Densities and Pressures}
\label{sec:ngcedges}

In Table \ref{tab:n7618ratios},
we present the effective density, temperature and pressure ratios
across the edges identified in the X-ray surface brightness profiles shown
in Figure \ref{fig:n7618profs} and listed in Table \ref{tab:n7618edges}.
For the Head and West Wing edges, the inferred effective pressure
ratios are close to $1$ (pressure equilibrium), as expected for
merger-induced sloshing, and the shape of the profile is also typical of
that expected for this model. However, the fitted density ratios (and
thus the effective pressure ratio across the edge) for
the outer plume edge and the eastern wing edge are unphysical (see
Plume2,3 and East Wing1,2, respectively, in Table \ref{tab:n7618ratios}). 
For $r<23$\,kpc, as one moves inside the  East Wing edge, the profile 
may average
over multiple features, such as part of the plume or an internal, sloshing
spiral similar to that seen in the West Wing. 
Fitting a single edge jump to 
multiple features would lead to an overestimate of the jump and thus
an unphysical estimate for the pressure ratio across the East Wing's
outer edge. The plume profile in Figure \ref{fig:n7618profs}
clearly shows a sharp rise at $r\sim 30$\,kpc merging smoothly into a
steeply rising profile with decreasing radius between the
outer and inner edges. The change in the surface brightness at the 
plume region's outer edge is a factor of  $2-3$, such that one would expect a 
physical density ratio of order $1.4-1.7$ and effective pressures consistent 
with pressure equilibrium.  However, from Figure \ref{fig:ngcanalysis} the
Plume2 region between the outer and inner plume edges contains 
excess emission from a filament that skews the shape of the profile from 
that expected by the simple effective density model used in the 
fitting algorithm.  This excess emission masks the characteristic 
flattening of the surface brightness profile inside the edge 
expected by the model.

\begin{deluxetable}{ccccc}
\tablewidth{0pc}
\tablecaption{NGC\,7618 Edge Analysis}\label{tab:n7618ratios}
\tablehead{ \colhead{Profile} & \colhead{$r_e$} &
  \colhead{$n_i/n_o$} & \colhead{$T_i/T_o$} & \colhead{$p_i/p_o$} \\
   & (kpc) & & &  }
\startdata
Head1,2 &$20.15$  &$1.70^{+0.28}_{-0.51}$  &$0.78^{+0.11}_{-0.07}$ 
 & $1.33^{+0.46}_{-0.48}$\\
Plume1,2 &$17.02$  &$1.58^{+0.19}_{-0.13}$  &$0.90^{+0.08}_{-0.12}$  
  & $1.42^{+0.31}_{-0.29}$\\
Plume2,3 &$29.65$  &$3.30^{+0.36}_{-0.28}$  &$0.79^{+0.16}_{-0.12}$  
&  $2.59^{+0.89}_{-0.61}$  \\
West Wing1,2  &$25.8$ &$1.42^{+0.15}_{-0.26}$
&$1.11^{+0.09}_{-0.10}$  &$1.57^{+0.31}_{-0.40}$  \\
West Wing2,3&$46.2$& $2.26^{+0.13}_{-0.17}$& $0.64^{+0.11}_{-0.09}$
&$1.45^{+0.32}_{-0.29}$ \\
East Wing1,2   &$23.3$  &$2.59^{+0.32}_{-0.24}$
&$0.85^{+0.09}_{-0.07}$
 &$2.20^{+0.45}_{-0.37}$  \\
\enddata
\tablecomments{The numbers (i, o) following the region 
  name indicate the spectral region (inside, outside) the corresponding 
  edge, respectively. 
  Uncertainites for derived values assume  extremes in the $90\%$ CL 
  uncertainties for measured properties.     }
\end{deluxetable}

\section{Discussion and Comparison to Simulations}
\label{sec:discuss}

Simulations by Roediger \etal (2015a, 2015b) and 
Sheardown \etal (2019) demonstrate that the origins of the asymmetric gas 
features and inferred gas flow patterns, i.e. the presence of shear 
flows or turbulence,  depend critically on the age and orbital
parameters of the merger. It is essential to determine whether the observed 
asymmetries are produced through hydrostatic instabilities or by other means 
before using their properties to constrain the
microphysics of the IGM. These simulations show that gas tails produced during 
merging activity  belong to two basic classes, ram-pressure tails or 
slingshot tails, based on how they are formed. Thus, the morphology and 
spectral properties of these tails are powerful tools to distinguish the 
orbital age of the merger. 

Ram-pressure tails are formed on the approach to perigee, as merging 
cores fall through IGM gas of increasing pressure.  IGM gas flows
around the cores are well described as flows around a
blunt object, with an upstream stagnation point at the leading cold 
front edge, and a thermal pressure ratio between the gas pressure at 
the stagnation point and in the undisturbed free stream region. 
The  ratio of these two thermal pressures can be used 
to infer the relative velocity between the core and the IGM 
(Vikhlinin \etal 2001). 
Strong shear flows around the core induce KHIs 
that create rolls, gas wings and boxy features and also strip the outer 
layers of gas from the merging core to form a 
tail (Roediger \etal 2012a, 2015a). 
The ram-pressure tail consists of three parts: a {\em remnant tail} closest to 
the merging core that is  composed of gas still gravitationally bound to 
the galaxy, a {\em dead water} region composed of stripped gas
that has little or no velocity relative to the merging core, and 
a {\em far wake} of stripped gas flowing away (downstream) from the core 
and mixing  with the IGM (Roediger \etal 2015a). 
Since mixing of the wake gas with the IGM can be strongly inhibited by 
IGM gas viscosity or the presence of magnetic fields, the observed length
and spectral properties of ram-pressure tails place important constraints on 
these properties (e.g. NGC\,1404 Su \etal 2017a, 2017b; 
NGC\,4552, Kraft \etal 2017).  While a ram-pressure  
tail is expected to point roughly opposite to the direction of infall of 
its merging core, the tail can be bent by bulk motions (winds) in the IGM, 
 caused by a previous encounter of the pair, a previous merger, or 
AGN outbursts. Thus, the observed ram-pressure tail morphology is
 a product of the entire merger history of the system and can be 
complex (Eckert \etal 2017; Kraft \etal 2017).   

{\em Slingshot} tails are formed as the merging cores move past each other at 
perigee and travel through IGM gas of decreasing pressure as the distance 
between the cores increases (Hallman \& Markevitch 2004; 
Ascasibar \& Markevitch 2006; Markevitch \& Vikhlinin 2007; 
Poole \etal 2006; ZuHone 2011). Simulations show that, as the cores near the 
apogee of their orbits, the remnant tails, composed of gas still
gravitationally 
bound to their parent system, display complicated morphologies and
temperature structures that depend strongly on the impact parameter of
the merger. For large impact parameters the tails form sweeping
arc-shaped structures. For small impact parameters, the tail gas
overruns the parent core producing a turbulent, irregularly shaped
secondary atmosphere about the parent before evolving into a
cone-shaped tail pointing away from the merger 
partner (Sheardown \etal 2019).  

\begin{figure*}[htb!]
\begin{center}
\includegraphics[width=4in,angle=0]{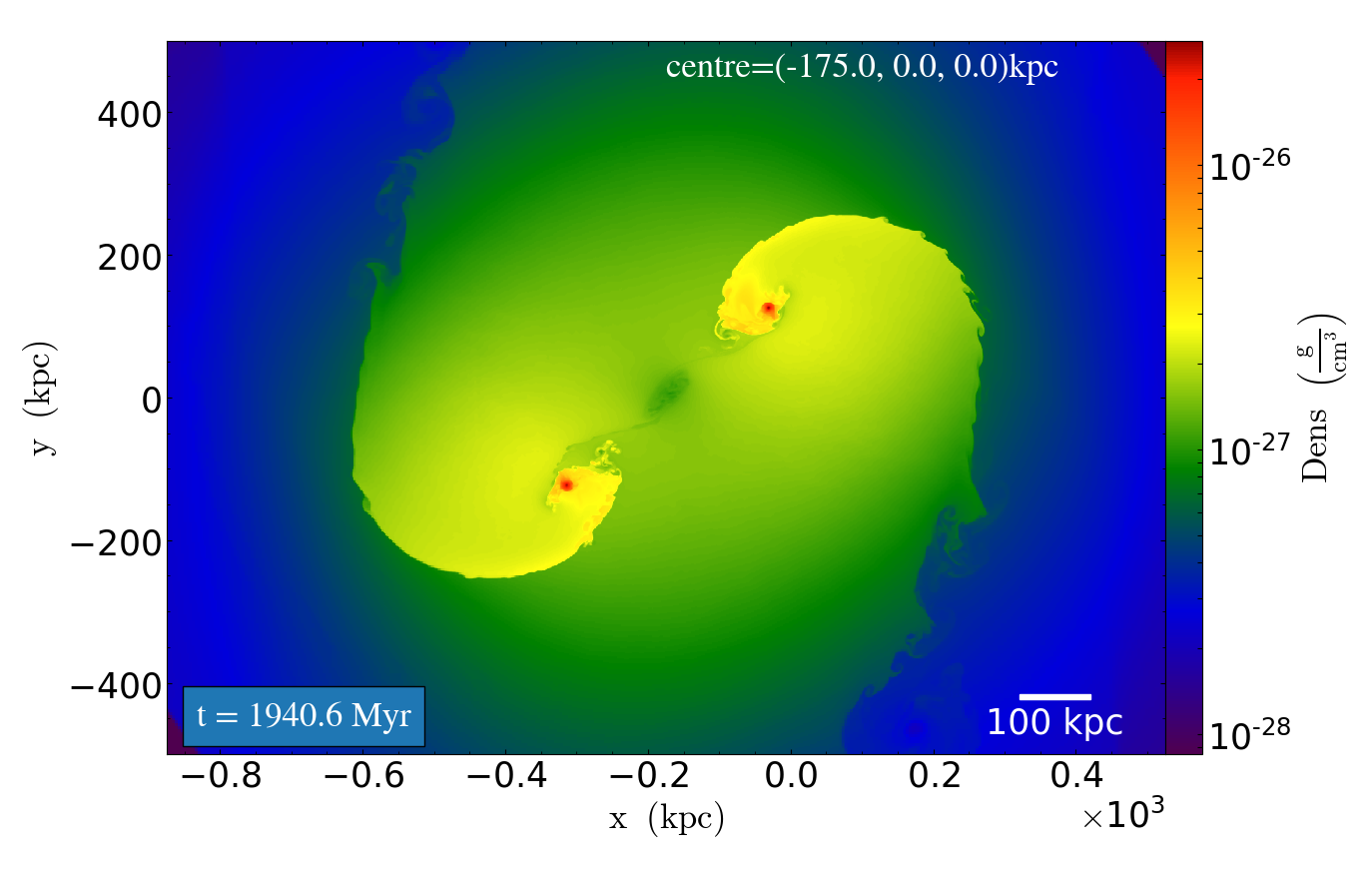}
\includegraphics[width=4in,angle=0]{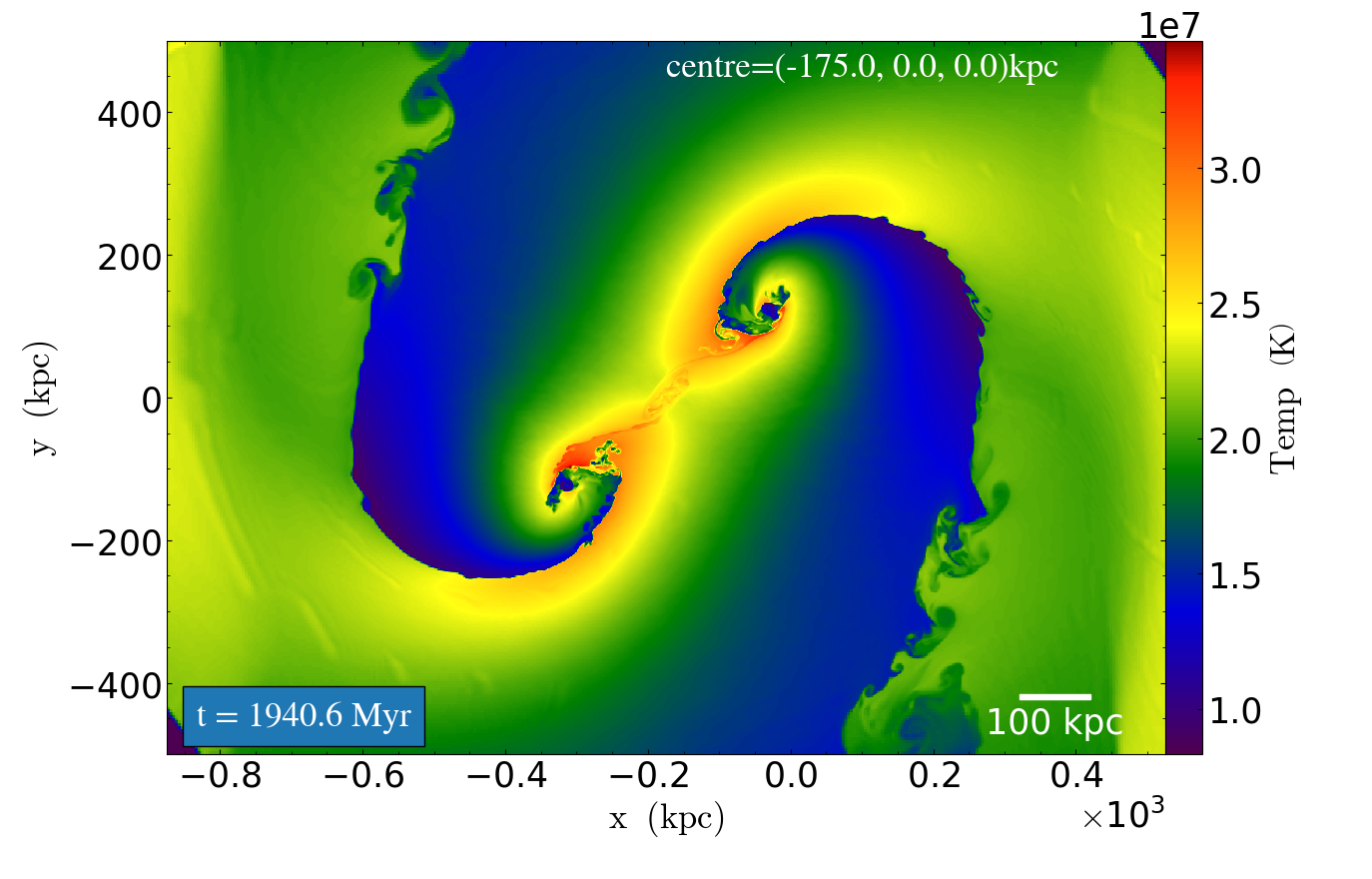}
\includegraphics[width=4in,angle=0]{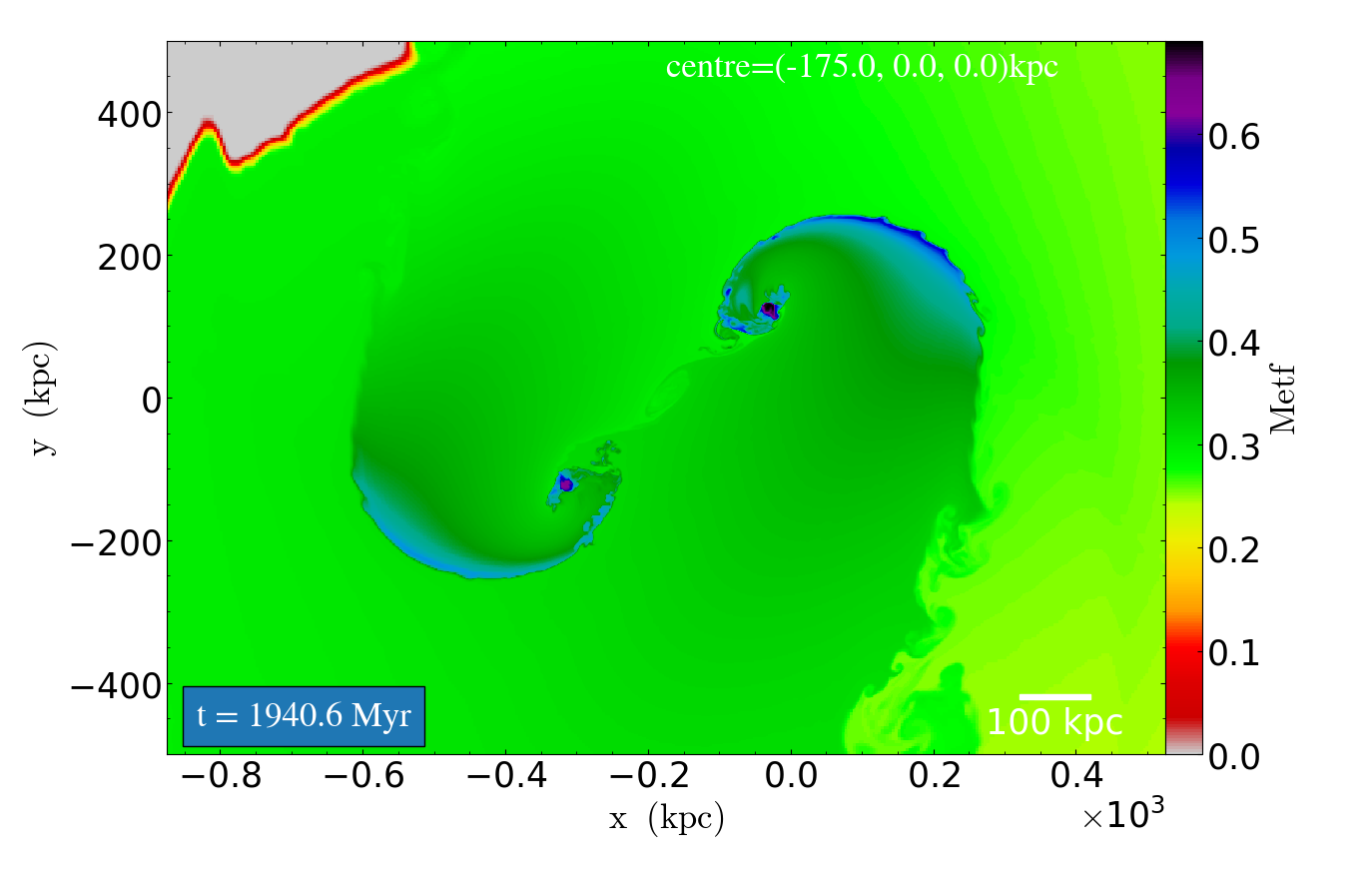}
\caption{ Slices at $z=0$ from a numerical simulation of the merger of
  two equal-mass galaxy groups with masses roughly equal to those of
  the NGC\,7618 and UGC\,12491 galaxy groups showing hot gas density
  ({\it upper}), temperature ({\it middle}), and metallicity 
  ({\it lower}) as the galaxy groups approach apogee after one
  pericenter passage.  }
\label{fig:comparesims}
\end{center}
\end{figure*}

In Figure \ref{fig:comparesims} we show gas density, temperature and
metallicity maps, 
derived from a numerical simulation of the merger of two equal mass galaxy 
groups at a merger age of $1.94$\,Gyr, as the galaxy groups approach apogee 
in their orbits after one previous pericenter passage (Sheardown \etal
2019). The mass of each simulated galaxy group in the merger was
$6\times10^{13}\Ms$. Thus, the total mass of the simulated merger is a
factor $3$ larger than the virial mass measured by Crook \etal (2007,
2008) for the HCD 1239 galaxy group containing NGC\,7618 and
UGC\,12491. Initial density profiles for the
groups were constructed as for the Fornax Cluster in Sheardown \etal
(2018). The initial separation was taken to be $1500$\,kpc, twice the
virial radius of each group so that initially the gas atmospheres of
each merger partner do not overlap. The initial tangential velocity
between the groups was chosen such that the impact parameter at the first
pericenter passage ($265$\,kpc) was large enough to produce the
arc-shaped slingshot tail morphologies seen in the observation. A
simulation run with smaller tangential velocity did not produce the 
observed well-defined arc-shaped tails, 
but rather resembled the overrun tail morphology discussed in Sheardown
\etal (2019). The gas was assumed to have zero viscosity. The maps in 
Figure \ref{fig:comparesims} were constructed by taking slices through
the corresponding simulation data cubes at $z=0$. 

\begin{figure*}[htb!]
\begin{center}
\includegraphics[width=6in,angle=0]{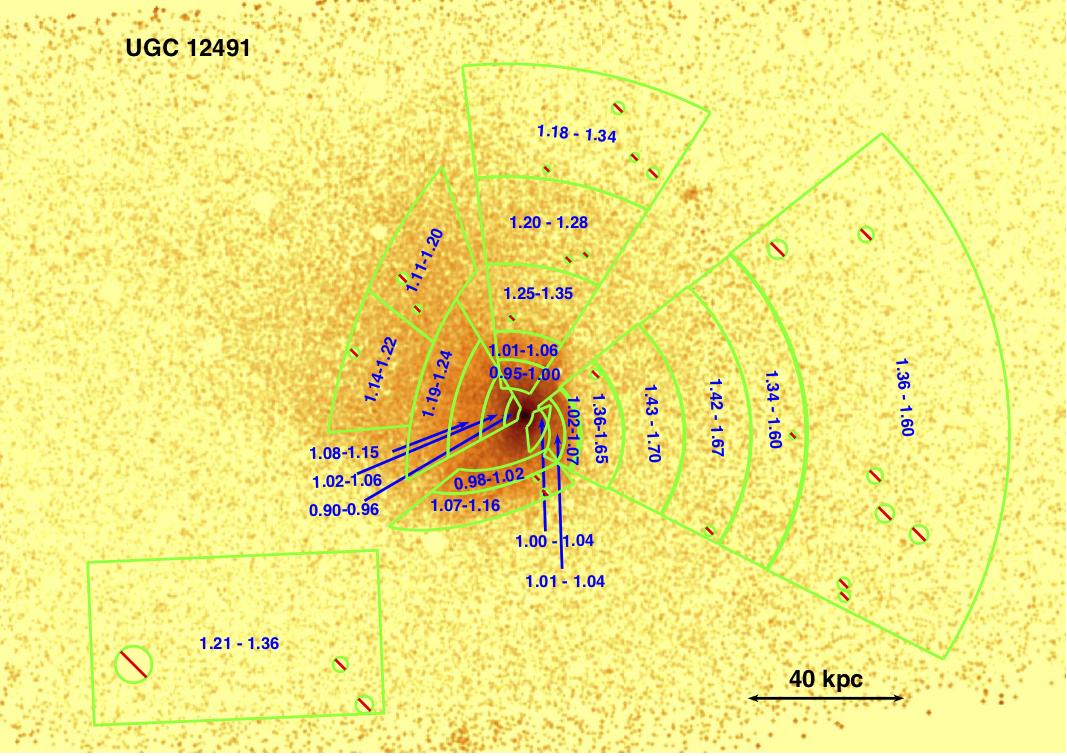}
\includegraphics[width=6in, angle=0]{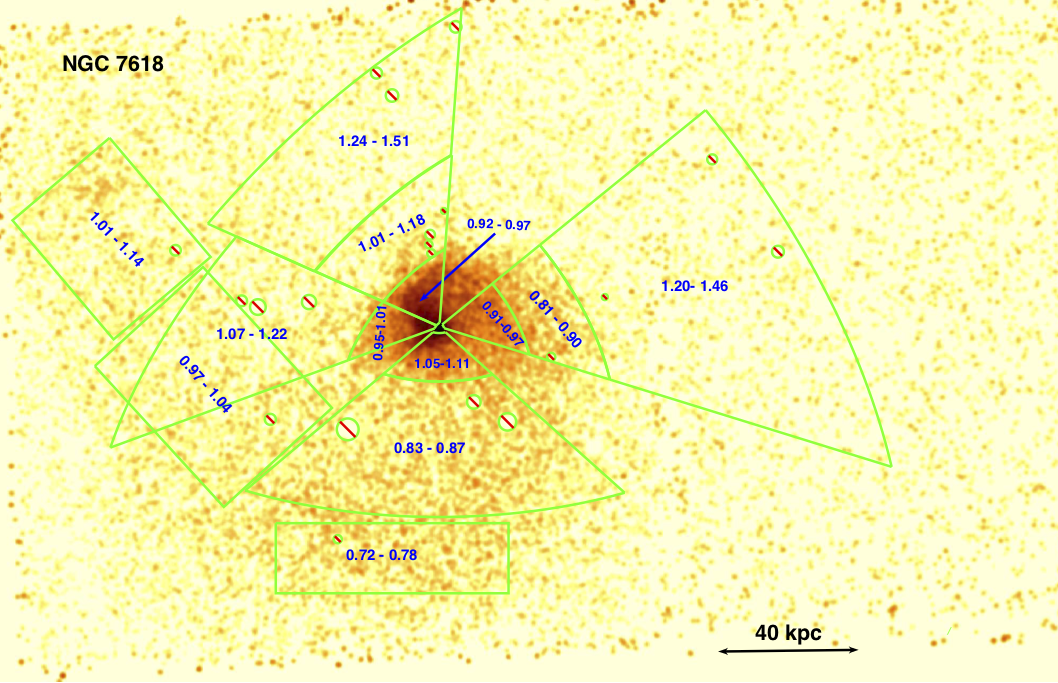}
\caption{APEC spectral model $90\%$ CL temperature ranges in keV units 
superposed on 
$0.5-2$\,keV {\it Chandra} X-ray  surface brightness images of UGC\,12491 
({\it upper panel}, see Table \protect\ref{tab:n7618spec} 
and Fig. \protect\ref{fig:ugcanalysis}) and NGC\,7618 ({\it lower panel}, 
see Table \protect\ref{tab:ugc12491spec}
and Fig. \protect\ref{fig:ngcanalysis}). The scale 
$40\,{\rm kpc} = 1.85$\,arcmin in each figure.}
\label{fig:obscomparesim}
\end{center}
\end{figure*}

\begin{figure*}[htb!]
\begin{center}
\includegraphics[width=6in,angle=0]{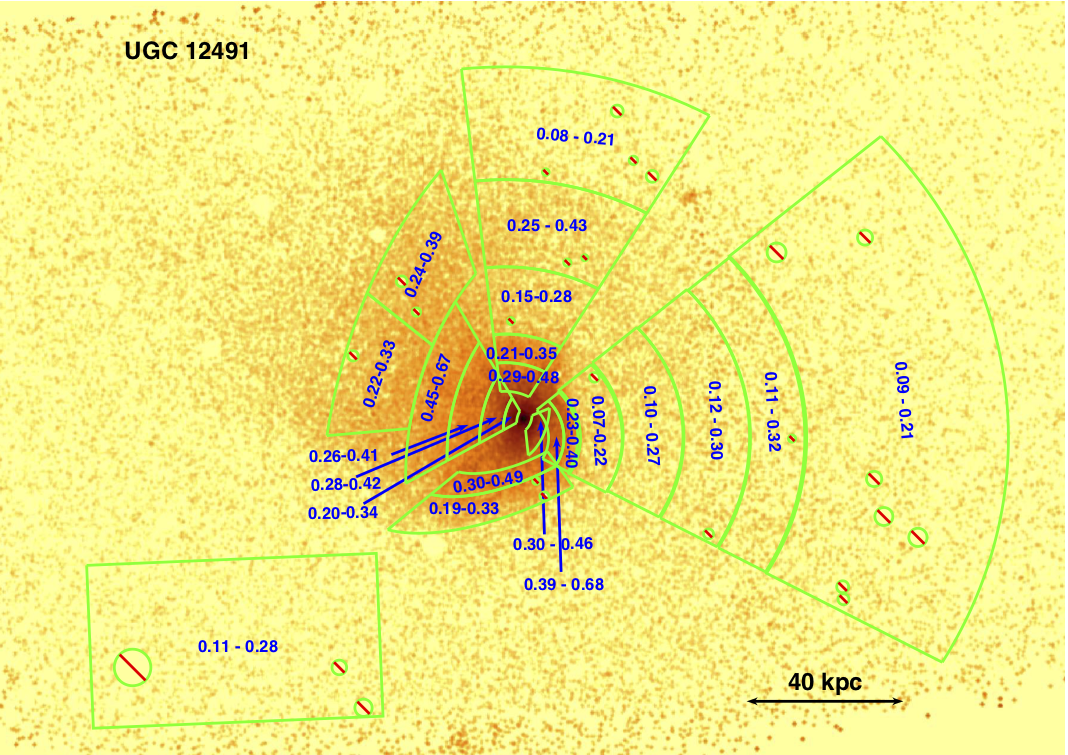}
\includegraphics[width=6in,angle=0]{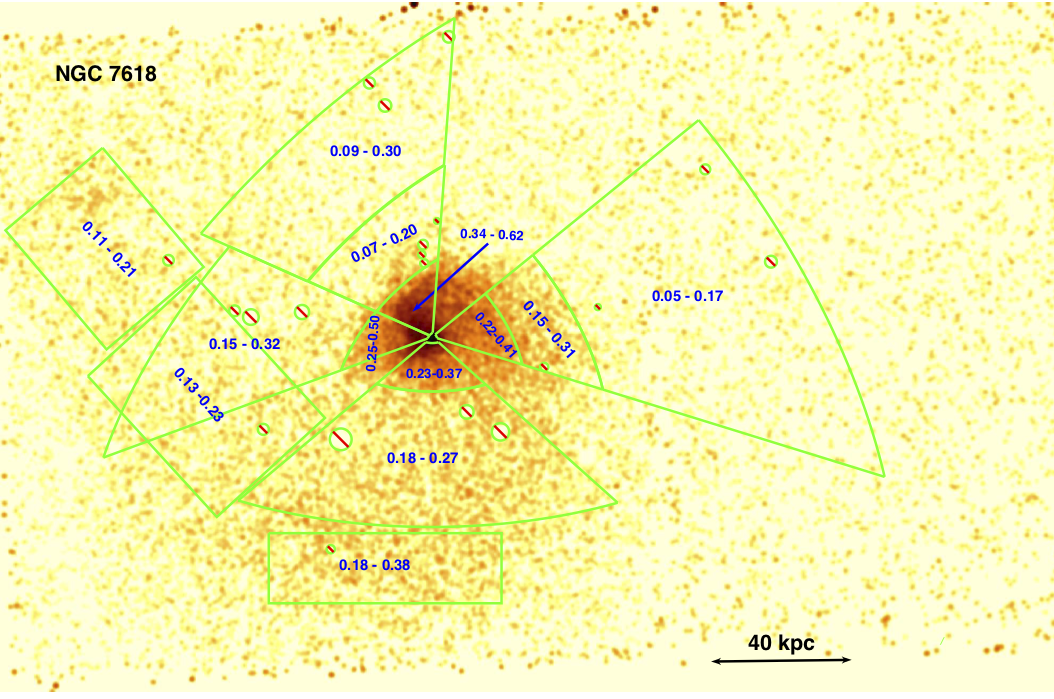}
\caption{APEC spectral model $90\%$ CL metallicity ranges in $\Zs$
  units superposed on the $0.5-2$\,keV {\it Chandra} X-ray  surface 
brightness images of UGC\,12491 ({\it upper panel}, 
see Table \protect\ref{tab:n7618spec} 
and Fig. \protect\ref{fig:ugcanalysis}) and NGC\,7618 ({\it lower panel}, 
see Table \protect\ref{tab:ugc12491spec}
and Fig. \protect\ref{fig:ngcanalysis}). The scale 
$40\,{\rm kpc} = 1.85$\,arcmin in each figure.}
\label{fig:obscomparemetals}
\end{center}
\end{figure*}

In Figures \ref{fig:obscomparesim} and \ref{fig:obscomparemetals}, 
we show the  $90\%$\,CL intervals for the hot gas temperatures 
and metallicities, respectively,  from the best fit APEC spectral 
models listed in Tables \ref{tab:ugc12491spec} and \ref{tab:n7618spec} 
 superposed on the $0.5-2$\,keV {\it Chandra} 
X-ray surface brightness images for UGC\,12491 and NGC\,7618 from
Figures \ref{fig:ugcanalysis} and  \ref{fig:ngcanalysis}, respectively. 
Figure \ref{fig:comparesims}, derived from the simulations,  and
Figures \ref{fig:obscomparesim} and \ref{fig:obscomparemetals},
derived from the observational data, are related, but not
quantitatively comparable. X-ray surface brightness, shown in 
the  observational images,
is proportional to the square of the electron density integrated along 
the LOS, while the simulation gas density map, 
constructed from a simulation slice, displays the gas density at 
a given LOS coordinate (here $z=0$). Similarly, the APEC spectral 
model temperature and metallicity $90\%$\,CL intervals superposed 
on the galaxy group X-ray images in Figures  \ref{fig:obscomparesim} and
\ref{fig:obscomparemetals} are projected (average) gas temperatures
and metallicities for all X-ray emitting gas along the LOS, while the 
corresponding simulated maps in Figure \ref{fig:comparesims} display 
a predicted gas temperature and metallicity for the specific LOS
coordinate. Despite these differences, the qualitative similarities
between the gas morphology and properties seen in maps derived from the 
numerical simulations
and  the related  maps  
derived from our {\em Chandra} X-ray data, are striking.

\subsection{Density and Surface Brightness Comparisons}
\label{sec:denscomp}

Both Figure \ref{fig:obscomparesim} and Figure \ref{fig:merge}
show arc-shaped tails bending sharply toward and/or wrapping around
the dominant merging galaxies. The arc-shaped morphology coupled with
the tight winding of the tails are 
signatures for slingshot tails formed from the merger of two galaxy
groups with a large impact parameter. 
Coupled with the small relative radial velocity between the two galaxies, 
the tail morphology strongly indicates that NCG\,7618 and UGC\,12491 are 
the dominant galaxies in two  equal-mass merging galaxy groups that are near 
apogee in their orbits. Comparison with the simulations of
Sheardown \etal (2019)
suggests a merger age of $\sim 1.94$\,Gyr with at least one previous pericenter 
passage. We compare our  high-resolution {\it Chandra} images of
UGC\,12491 and NGC\,7618 
in Figures \ref{fig:ugcanalysis} and \ref{fig:ngcanalysis},
respectively,
to the 
simulation density slice in the upper panel of Figure \ref{fig:comparesims} to 
study the fine structure in the gas. We first compare the detailed
morphology of the NGC\,7618 and UGC\,12491 merging cores to that seen in the
simulated galaxies, and then compare the morphology of the tails. 

\subsubsection{Density Morphology in the Remnant Cores}
\label{sec:corecompare} 

The simulated galaxy  to the southeast in Figure \ref{fig:comparesims} shows 
features in the density slice similar to the observed X-ray surface brightness 
features in NGC\,7618 shown in Figure \ref{fig:ngcanalysis} and the
lower panel of Figure \ref{fig:obscomparesim}.
The nose and northern edge in both
the southeastern simulated galaxy and NGC\,7618 
show low-amplitude rolls or scallops, likely caused by hydrodynamic 
KHIs. Such KHIs are  expected in inviscid gas.
NGC\,7618 and the southeastern simulated galaxy have a broad western
wing with a boxy edge to the north. The bright plume in NGC\,7618 is
not as prominent in the southeastern simulated galaxy. 
Filamentary features are found in both the {\it Chandra} X-ray image
of NGC\,7618 and in the simulation slice, but in contrast to the
observed filament that emerges from 
the nose of NGC\,7618 near the plume, the filaments found in the
southeastern simulated galaxy are north of the western wing along the
line connecting the remnant cores. However,
these turbulence-related plumes and filaments are thought to be
transient features and 
exact matching of length and location should not be expected.  

The simulated galaxy to the northwest in Figure \ref{fig:comparesims}
displays features similar to those observed in our high resolution
{\it Chandra} X-ray surface brightness images of UGC\,12491 in
Figure \ref{fig:ugcanalysis}. UGC\,12491 and the 
northwestern simulated galaxy each have a boxy North Wing feature.
However, the observed 
edge in the North Wing of UGC\,12491 is smoother than that in the
northwestern simulated galaxy. A higher-density plume, directed from
the galaxy's center toward its nose, is 
seen in the northwestern simulated galaxy.  The plume is likely pushing gas 
outward, making the nose edge of the simulated galaxy  irregular on scales of 
$\sim 20$\,kpc.  This plume in the northwestern simulated galaxy is
much more like the 
plume observed in NGC\,7618. In UGC\,12491 the central elongated dense
gas is oriented 
nearly parallel to the nose, like a sloshing edge, and the observed
irregularities at the nose edge are an order of magnitude smaller.  
 
\subsubsection{Gas Morphology of the Slingshot Tails}
\label{sec:tailcompare}

The observed tails in the {\it Chandra} X-ray surface brightness
imagesof NGC\,7618 and 
UGC\,12491 are remarkably similar in morphology to the slingshot tails
found in the 
simulation density slice of the merging southeastern and northwestern galaxies, 
respectively, at apogee. 
In the southeastern simulated galaxy, the tail extends southward from 
its western wing.  This supports the hypothesis,
presented in \S\ref{sec:n7618spectra}, 
that the outer part of the western wing in NGC\,7618 (the West Wing2
region) is tail gas before 
the tail is seen to wind southward around the galaxy. In both NGC\,7618 and the 
southeastern simulated galaxy the tails extend to the
south for $\sim 100$\,kpc before 
turning to the north and northeast. The inner edge of each tail winds close 
to its parent galaxy's eastern wing. 

The simulated tails in Figure \ref{fig:comparesims} are expansive with widths 
$\gtrsim 200$\,kpc after the tails  bend back toward their parent
cores.  The outer edges of the tails remain well defined for $>
400$\,kpc along the length of the tail, while the gas 
density decreases by a factor $\sim 2$. Note that such an extent is
larger than the $8.3^{\prime}\times 8.3^{\prime}$
($180\,{\rm kpc} \times 180$\,kpc) ACIS S3 field of view (FOV) 
at the redshift of the NGC\,7618/UGC\,12491 system. The outer edge of
NGC\,7618's tail remains well defined along the observable length of
the tail, i.e until the outer edge of the tail falls outside the FOV
of the combined observations (see Fig.\ref{fig:merge}). 
While this general morphology of NGC\,7618's tail is qualitatively
similar to the morphology of the tail in the southeastern simulated
galaxy, the observed width of the tail in NGC\,7618, measured east
from the remnant core is only half that found 
in the simulated tails. As pointed out by Sheardown \etal (2019), this
may be at least partly due to a projection effect. If the merger plane
is inclined relative to the LOS, 
the simulation tails appear narrower and more tightly wound 
(see Fig. 7 in Sheardown \etal 2019).  

The slingshot tail in the northwestern simulated galaxy 
in Figure \ref{fig:comparesims} has two higher-density streams that
make it appear bifurcated as it curves to the north after emerging
from the remnant core's south wing.
In Figures \ref{fig:merge} and  \ref{fig:ugcanalysis}, we can see the
same bifurcated morphology and bend to the north in the X-ray surface
brightness images of UGC\,12491's tail. The northwestern simulated
galaxy's tail is tightly wound to the west and south around the galaxy
core, such that the simulated gas density west of the core is higher
than that to the southwest along the line joining the merging
partners. A similar region of enhanced X-ray surface brightness
corresponding to higher density gas, that may be an extension of its tail,
 is found west of UGC\,12491 as shown in Figure \ref{fig:merge}. However, 
the outer extent of this higher density region to the west of
UGC\,12491 is more diffuse than the well-defined outer edge expected
for the tail, and the width of the region is  less 
than that expected from the simulation density slice.         
  
Low-amplitude KHI rolls are seen along the outer edges of the tails in
the southeastern and northwestern simulated galaxies. The amplitude of
these KHIs grow as the distance along the tail increases and the gas
density decreases. These KHIs at larger distances along the tails
would be faint and difficult to detect in our data.

\subsection{Temperature and Metallicity Comparisons}
\label{sec:tempcomp}

 The simulated temperature and metallicity maps in the middle and
 lower panels of Figure \ref{fig:comparesims} and the observed
 projected temperature and metallicity maps 
in Figures \ref{fig:obscomparesim} and \ref{fig:obscomparemetals} show
qualitatively similar, complex temperature and metal abundance 
structures. In the simulated temperature slice, gas heated by the
merger interaction fills the space between the two galaxies and
surrounds the cool galaxy gas cores, dense filaments, and extended
boxy features. This is also seen in the observational data. 
From the best-fit APEC spectral models listed in Table \ref{tab:ugc12491spec}
for UGC\,12491 and in Table \ref{tab:n7618spec} 
for NGC\,7618, the projected temperature for gas
filling the space between UGC\,12491 and NGC\,7618, 
measured in the South IGM and West Wing3 regions, respectively, is
higher ($\sim 1.2-1.3$\,keV) and the metallicity is 
lower ($0.1-0.2\,\Zs$) than in the cool ($0.9-1.0$\,keV) merger cores,
boxy features, and filaments that are composed of higher metallicity
($\sim 0.3-0.4\,\Zs$) galaxy gas
(see the Nose1,2, North Wing1,2, Inner W1,2, South Wing1 regions
in the upper 
panels of Figs. \ref{fig:obscomparesim} and \ref{fig:obscomparemetals}
and in Table \ref{tab:ugc12491spec}
for UGC\,12491 and the  Head1, Plume1, East Wing1, and West Wing1
regions in the lower panels 
of Figures  \ref{fig:obscomparesim} and \ref{fig:obscomparemetals} and 
in Table \ref{tab:n7618spec} for NGC\,7618). 

The highest temperature gas in the simulation lies just outside the
leading edges of the cool merger cores. In the southeastern simulated
galaxy, the highest temperature gas extends $\sim 50$\,kpc to the
north of the core, while the highest temperature gas 
outside the leading edge of the northwestern simulated galaxy is
confined to a narrow ($< 10$\,kpc) region adjacent to the edge. Since
the simulated galaxy groups are of equal mass, this difference in the
morphology of the regions is likely an artifact 
of the viewing angle of the simulation slice relative to the orbital
plane of the merger.   The highest temperature ($1.4-1.7$\,keV) gas
measured in the observed NGC\,7618/UGC\,12491 merger 
is found west of the leading edge of UGC\,12491 (see the Nose3-7 regions 
in Table \ref{tab:ugc12491spec} and in Figure
\ref{fig:obscomparesim}).  This large high-temperature region is 
more like the high-temperature region found north of the leading edge
of the southeastern simulated galaxy rather than the northwestern
 simulated galaxy, as might be expected from comparisons of gas
 morphology in \S\ref{sec:denscomp}. 
However, the high-temperature region observed west of 
UGC\,12491 extends twice as far ($\sim 100$\,kpc ) from the leading
edge as that of the highest temperature region in the southeastern
simulated galaxy. Also the observed X-ray surface brightness, and
thus gas density, west of UGC\,12491, shown in 
Figure \ref{fig:merge}, is higher than that to the south of UGC\,12491 between 
the merging cores, while in the southeastern simulated galaxy,
the highest temperature gas is found in a region of lower density than
that found elsewhere between the simulated cores. 
North of NGC\,7618, we do not observe a significantly higher
temperature region close to the leading edge, as one might expect
from the simulated temperature slice. If the 
high-temperature region is narrow or contaminated 
with cool filaments of galaxy gas, as is the case for the Plume2
region in NGC\,7618,  
the limited photon statistics available for spectral analysis in
NGC\,7618 would make such a high-temperature (lower-density) region
difficult to detect.  

In the simulated merger, streamers of hot gas, displaced from the 
highest temperature regions in front of the leading edges of each galaxy by the 
close encounter, form a high-temperature, low-density 
bridge midway along the line connecting the two galaxy cores.
As shown in Figure \ref{fig:comparesims}, this feature persists in the
simulations even as the merging partners approach apogee. Since that
region falls far from the aimpoints of 
our observations, it is not surprising that we find no evidence for
such a high-temperature bridge in our data.

\subsubsection{Temperature and Abundance Comparisons in the Tails}
\label{sec:tailtempcomp}

As discussed in \S\ref{sec:tailcompare} and seen in
Figure \ref{fig:comparesims}, 
the tails in the simulation slices emerge 
from the west wing for the southeastern galaxy and south wing for the 
northwestern galaxy. Hot gas with temperatures equal to that of gas between the 
merger cores follows the outer edge of the simulated tails from their apparent 
origin in the wings until the wide bend around their parent cores 
(see the middle panel of Fig. \ref{fig:comparesims}). The  
$1.32^{+0.14}_{-0.12}$\,keV temperature of gas measured in the
West Wing3 region in NGC\,7618 just outside the emerging tail in
West Wing2 is consistent with this prediction. 
  
The simulation temperature slice shows that in and near the western wing in the 
southeastern galaxy and the south wing in the northwestern galaxy, 
the distribution of gas within the tails is composed of smaller, irregular 
regions of different temperature gas. The temperatures in these
irregular regions
are intermediate between the temperature of the gas in the cool merger 
cores and the  high-temperature gas between the merger cores and 
 in front of their leading edges. Physically this may be the 
slingshot tail gas pushing past and being disturbed by the 
slowing remnant core as the core approaches apogee. 
In Figure \ref{fig:comparesims}, regions 
of multi-temperature gas within the tail 
are  most apparent in the northwestern simulated galaxy, where these regions 
of different densities, temperatures and abundances cause the tail to appear 
bifurcated. As discussed in \S\ref{sec:tailcompare},
the {\it Chandra} X-ray image 
of the observed tail in UGC\,12491 shows a similar morphology. 
In UGC\,12491 the intermediate temperatures ($\sim 1.2$\,keV) measured 
in the North Tail, South Tail, and Inner Tail E4 regions in UGC\,12491
compared to $0.9-1.0$\,keV in the galaxy and the $1.3$\,keV gas
between the cores (the South IGM region in  Table \ref{tab:ugc12491spec}) 
are qualitatively consistent with what one would expect for an average over 
such  multi-temperature gas. The metal abundance in the Inner Tail E4 
 region is also significantly higher than that measured in the other tail
regions (see the upper panel of Figure \ref{fig:obscomparemetals}). 
Although regions of higher metallicity gas are expected in the tail, the lower 
panel of Figure \ref{fig:comparesims} shows that, in the simulation
metallicity slice, the regions with higher metal abundances are found
near the outer edges of the tail, while in UGC\,12491 the measured
high metallicity region is on the tail's inner edge.

Farther from the wings, as the tail curves sharply around the galaxy, 
the temperature of the gas in the simulation  
decreases smoothly across the width of the tail from higher temperatures 
for gas on the inner (most tightly wound) edge of 
the tail to lower temperatures for  gas on the outer edge of the tail. 
In the middle and lower panels of 
Figure \ref{fig:comparesims}, the lowest 
temperature, highest metallicity gas in each tail is comparable to the 
temperature and metallicity in the inner regions of the parent remnant  
core, and is found along the outer edge of the tail, along the final
wide bend as the tail changes direction back toward its parent galaxy. 
 In NGC\,7618 we are able to trace the tail as it changes 
direction from trailing south of West Wing2 through its region of maximum 
curvature until it winds northeast back towards NGC\,7618's core. 
As discussed in \S\ref{sec:n7618spectra}, the South Tail 1,2 and 
Southern Most Tail regions measure temperatures across the width of the
tail as it makes this change in  direction. The decreasing temperatures 
measured from near the galaxy ($1.07^{+0.04}_{-0.02}$\,keV in South Tail1) 
through the center of the tail ($0.83^{+0.02}_{-0.02}$\,keV in South Tail2)  
to the tail's outer edge ($0.75^{+0.03}_{-0.03}$\,keV in Southern Most Tail) 
are consistent with the temperature gradient expected 
in this region from the simulations.  After the tail has completed its
turn to the northeast, the gas temperatures measured in the  
Mid Tail and North Tail regions in NGC\,7618 average 
over the width of the tail. Thus the measured intermediate temperatures 
$\sim 1$\,keV are also qualitatively consistent with the simulation.  

\section{Conclusions and Future Work}
\label{sec:conclude}

In this paper we used {\it Chandra} X-ray observations to study
the properties of hot
gas in the major merger of the NGC\,7618 and UGC\,12491 galaxy groups,
and compared our 
findings to simulations of a merger of two equal mass galaxy groups
roughly matched to the total mass, density and temperature profiles
expected for the NGC\,7618 and UGC\,12491 galaxy groups. 
Gas in the simulations was assumed 
to have zero viscosity. 
X-ray spectra from our {\it Chandra} data are modeled using absorbed single temperature 
APEC models throughout this study. 

We find the following: 
\begin{enumerate}
\item{ From $0.5-2$\,keV {\em Chandra} X-ray surface brightness
    images, we identified
   edges, wing-like and boxy features, filaments and tightly wound tails
   in the hot gas of each galaxy group.}
\item{ X-ray surface brightness profiles across features of interest
    contain  multiple surface brightness discontinuities (edges) that
    are  well characterized using spherically symmetric power-law models
    with a density contact discontinuity at each edge. These edges
    are largely consistent with sloshing, as expected in merging
    systems.}
\item{Hot gas in the merger cores, boxy features, wings and filaments 
    have temperatures $\sim 1$\,kev and metallicities $\sim 0.3\,\Zs$.}
\item{Hot gas between NGC\,7618 and UGC\,12491 has a  higher 
   temperature ($\sim1.3$\,keV) and lower metal abundance 
   ($\sim 0.1-0.2\,\Zs$) than within the merger cores. The highest 
   temperature gas ($\sim 1.5-1.6$\,keV) is found west of UGC\,12491's 
   sharp edge.}
\item{Comparison of these observational
    results with simulations of an equal-mass merger of two galaxy
    groups roughly matched to NGC\,7618 and UGC\,12491 suggests that
    we are viewing the NGC\,7618 and UGC\,12491 groups at a merger age
    of $\sim 2$\,Gyr after one pericenter passage, when the galaxy
    groups are at the apogee of their orbits.}
\item{The striking qualitative similarities between these simulations
    using inviscid gas and our observations suggest that the
    observed hot gas in the NGC\,7618/UGC\,12491 merger has zero 
    viscosity.}
\item{The observed tails are arc shaped and tightly wound about their
    respective merging cores, consistent with slingshot tails formed
    after a pericenter passage at a large impact parameter.}
\item{Spectral fits across NGC\,7618's tail near maximum curvature show 
    that the gas temperature decreases from the inner to the outer edge of
    the tail, consistent with that expected from simulations.}
\end{enumerate}
 
While the qualitative similarities between the observations and
simulations are encouraging, differences still exist: 
\begin{enumerate}
\item{ The observed tails in Figure \ref{fig:merge} and
  Figures \ref{fig:ngcanalysis} and 
  \ref{fig:ugcanalysis} appear narrower than those in the simulated d
  ensity slice shown in 
  Figure \ref{fig:comparesims}. However, simulations  show that if the
  merger plane is not in the plane of the sky, i.e., the merger plane is 
rotated relative to the LOS, the simulated projected X-ray 
intensity of arc-shaped slingshot tails may appear narrower and more
tightly bound than if the merger plane is in the plane of the sky
(see Figure 7 in Sheardown  \etal 2019). It is
not clear whether the apparent narrow width of the observed slingshot
tails for NGC\,7618 and UGC\,12491 can be reproduced in these
simulations by simply varying the angle of the LOS relative to the
merger plane for projected observables, or is an 
observational artifact of our limitied statistics and ACIS-S FOV,
or may point to additional physics important to the merger not
included in existing simulations.}

\item{The highest temperature gas measured in our observations is
  located just west of the sharp ``Nose'' edge of UGC\,12491. 
  Simulations and our X-ray surface brightness image in
  Figure \ref{fig:merge} suggest that the increased density west of
  UGC\,12491 is an extension of  UGC\,12491's arc-shaped tail as it
  winds around its remnant core. However, simulations expect
  arc-shaped slingshot tails to be primarily composed of cool,
  unmixed galaxy gas, in contradiction to the high temperatures
  observed in that region.  Thus the origin of this hot gas is
  unclear. Could this gas be heated (shocked) IGM gas rather than
  an extension of the tail? This latter interpretation may be more
  consistent with the  measured high temperature in a region of
  higher gas density. }

\item{The NGC\,7618 and UGC\,12491 galaxy groups are of approximately
    the same mass. Simulations predict that the remnant parent cores
    develop tails of similar size and that appear symmetrical in
    projected X-ray intensity maps even if the merger plane is rotated
    relative to the LOS (Sheardown \etal 2019). However, NGC\,7618's
    tail does not appear as tightly wound about its parent core as
    that of the tail about UGC\,12491. This again may be a
    projection effect due to the orientation of the LOS relative to the plane of
    the merger. Alternatively, the observed asymmetry could possibly
    result from differences in the overall group potentials due to
    different ellipticities or size, despite similar masses,  or
    reflect different interaction histories of the groups prior to
    this merger that might, for example,  introduce different initial
    velocities of the groups prior to their first encounter. Such
    asymmetries would need to be reflected in the initial parameter
    space used to construct simulation models.}  
\end{enumerate}

Given their low redshift, the merger of NGC\,7618 and UGC\,12491 may
be the best example 
in the local universe of a major merger of two low mass galaxy groups
at apogee, and, thus,
is deserving of further study.  We note that the current simulations
were only designed 
to provide sensible initial dark matter density, IGM density and
temperature profiles for 
a major merger of two galaxy groups similar to that of NGC\,7618
and UGC\,12491. They were not
designed to quantitatively match the level of detail revealed by the current 
observations. Simulation studies that probe  the origins of the differences 
discussed above, as well as the prevalence and resilience of
turbulence-induced, transient  
features, such as the filaments, plumes and bifurcated tails seen
in these {\it Chandra} data, 
are interesting topics for future study.  

Similarly, improved observations are needed to 
inform and test the simulation results. The ideal instrument for
future observations should 
couple  large effective area at temperatures $\sim 1$\,keV with
large FOV to map the 
full extent of the tails, measure density, temperature and
metallicity distributions 
throughout the tails and in the IGM between and around the
dominant galaxies, and to identify
the faint, large scale KHI features predicted to form along a
slingshot tail's outer edge. 
Detailed studies of the smaller scale turbulence-induced features,
such as filaments or plumes,
 and edge-like sloshing features associated with the 
 merging cores would also require high ($\sim {\rm arcsecond}$)
 spatial resolution. 
 Such future simulation and observational studies will continue to
 sharpen our understanding of 
galaxy group merger dynamics and the microphysics  
of hot gas throughout the merging system  
during this important stage of structure formation.   
 
\section*{Acknowledgments}
\label{sec:ack}

This work was supported in part by NASA {\em Chandra} grants GO4-15082X, and 
by the Smithsonian Institution. J.T. W. gratefully 
acknowledges support from a Bill Davis SURF Fellowship. 
Data reduction and analysis were supported by the Chandra X-ray Center 
 CIAO software and calibration database.  
The NASA/IPAC Extragalactic Database (NED), which is operated by 
JPL/Caltech, under contract with NASA was used throughout.  
We thank M. Markevitch for the surface brightness profile construction 
and analysis routines and L. Lovisari for useful discussions.

\begin{appendix}

\section{Profile Fitting Across Multiple Edges}
\label{sec:proffitproc}

In this section we present the spherically symmetric single- and
multiple-edge density models
used to fit X-ray surface brightness profiles for NGC\,7618 and
UGC\,12491. We caution the reader that for regions outside the galaxy
where the features show a complex morphology, spherical symmetry may
not be a good approximation and these simple models
may only provide a parameterization of the structure rather than
a measurement of the physical, electron density profile. By
parameterizing the complex structure in simulation data in the same
way as the observational data, one can  use the simulation to infer   
the unprojected three dimensional electron density distribution 
of the hot gas.

For profiles with a single surface brightness
discontinuity (edge) located at a radius $r_e$ and with jump $j_e$,
the electron gas density model is given, moving inward from large
radii, by  
\begin{equation}
  \begin{array}{l l}
      \tilde{n}_o(r>r_e) = & A\left(\frac{r}{r_e}\right)^{\alpha_o} \\ 
     \tilde{n}_i(r<r_e) = & Aj_{e}\left(\frac{r}{r_{e}}\right)^{\alpha_i} 
  \end{array}
\label{eq:oneedge}
\end{equation}
with $A$ the normalization, $\alpha_o$( $\alpha_i$) the density power
law indices outside (inside) the edge, $r_e$ the location of the edge,
and $j_e = \tilde{n}_i(r=r_e)/\tilde{n}_o(r=r_e)$ the jump (Vikhlinin
\etal 2001).  $\tilde{n}_x$  is
a fit variable proportional to the electron density $n_x$ and defined as 
$\tilde{n}_x = \sqrt{\Lambda_x}n_x$,$x={ o,i}$ for outside and inside the edge 
respectively.  The surface brightness profile is fit by integrating
the emissivity per unit area, $\tilde{n}^2(r)$, along the LOS for each
$r$ using a multivariate $\chi^2$ minimization algorithm, allowing the edge
location $r_e$, jump $j_e$ and power law indices $\alpha_o$ and
$\alpha_i$ to vary. 

For multiple edges we repeat this modeling for each edge requiring the
power-law index to be the same between successive edges. However, we
do not have sufficient statistics to fit all the variables at once, so
we proceed iteratively from large radius to small. For two edges, we
first fit the outermost edge with Eq. \ref{eq:oneedge}. Note that 
the power-law slope $\alpha_2$ between the outer and inner edges depends 
sensitively on the fit to the inner edge and it, in turn, affects 
the best-fit value for the jump at the outer edge. Thus we freeze only the
outer power-law slope $\alpha_3$ and outer edge 
location $r_{eo}$ from the first fit, but allow the outer jump 
$j_o$, power-law slope between the edges $\alpha_2$, 
the inner edge location $r_{ei}$ and jump $j_i$, and the inner
power-law index $\alpha_1$ to vary. We then refit the data from large to
small radii using Eq. \ref{eq:twoedge}.   

\begin{equation}
  \begin{array}{l l}
     \tilde{n}(r>r_{eo}) = & A\left(\frac{r}{r_{eo}}\right )^{\alpha_3}\\
     \tilde{n}(r_{ei} < r < r_{eo}) = & A j_o\left(\frac{r_{ei}}{r_{eo}}\right )^{\alpha_2}\left(\frac{r}{r_{ei}}\right )^{\alpha_2} \\ 
     \tilde{n}(r < r_{ei})= & Aj_ij_o\left (\frac{r_{ei}}{r_{eo}}
       \right )^{\alpha_2}\left (\frac{r}{r_{ei}}\right )^{\alpha_1}   \\
  \end{array}
\label{eq:twoedge}
\end{equation}

This procedure is then iterated for additional edges. For these data
the maximum number of edges identified is three. Equations
\ref{eq:oneedge} and \ref{eq:twoedge} are used to fit the outer and
middle edges. The full set of model parameters are power-law indices 
$\alpha_i, i=1,4$ with increasing integer for increasing radius, as in 
Tables \ref{tab:ugc12491alpha} and \ref{tab:n7618alpha}, three
edge locations and three associated jumps ($r_{ex}$ and $j_x, x={o,m,i}$ 
 for outer, middle and inner edges, respectively. Fitting iteratively from 
large radii to small, $r_{eo}$ and $\alpha_4$ are fixed by the first
iteration using Eq. \ref{eq:oneedge} to fit the outer edge, $j_o, r_{em}, 
\alpha_3$ by the second iteration fitting  the outer two edges using 
Eq. \ref{eq:twoedge}. The final iteration then determines the
remaining parameters ($\alpha_1$,$\alpha_2$, $r_{ei}$, $j_i$, $j_m$ ).  
 
\section{Definitions of Spectral Regions}
\label{sec:specdefs}

In Tables \ref{tab:ugc12491bounding} and \ref{tab:ugc12491specregs} we
define in detail the spectral regions shown in the lower panel of Figure  
\ref{fig:ugcanalysis} for the UGC\,12491 galaxy group. 
All of the spectral regions other than the rectangular (box) region
used to determine the temperature and abundance of the IGM to the
south of UGC\,12491 are elliptical panda regions, confined to the 
sectors also listed in Table \ref{tab:ugc12491specregs}.  To measure 
the properties of the hot gas across the X-ray surface brightness
discontinuities as cleanly as possible, 
we define a bounding ellipse that matches the shape of the 
surface brightness edges  within the sector of interest.  The bounding
ellipses for the regions shown in Figure \ref{fig:ugcanalysis} are 
listed in Table \ref{tab:ugc12491bounding}. Note that the bounding ellipses
for the tail regions and the Inner W region  are each 
centered on the X-ray peak of UGC\,12491, while the bounding ellipses
for the Nose and North Wing  are offset from that peak.  
In Table \ref{tab:ugc12491specregs} we list the (semimajor, semiminor) radii
for the inner and outer ellipses, each concentric to the bounding
ellipse in Table \ref{tab:ugc12491bounding}, and that, when
constrained to lie in the corresponding sector given in columns ($4$)
and ($5$) of Table \ref{tab:ugc12491specregs}, complete the definition
of the spectral regions shown in Figure \ref{fig:ugcanalysis}. We also
define the rectangular region used to measure the properties of the
group gas south of UGC\,12491 in the direction of the merging partner
NGC\,7618.

In Tables \ref{tab:n7618bounding} and \ref{tab:n7618specregs} we
present similar data to define the spectral regions for NGC\,7618
shown in the lower panels of Figure \ref{fig:ngcanalysis}. In 
Table \ref{tab:n7618bounding} we list the bounding ellipses used to
construct the X-ray surface brightness profiles and the elliptical
panda spectral regions. All of the bounding ellipses are centered at
the X-ray/optical center ($\alpha=23^{h}19^{m}47.22^{s}$, 
$\delta=+42^{\circ}51^{'}09.5\as$) of NGC\,7618. As in 
Table \ref{tab:ugc12491specregs}, the definition of the region is 
completed by listing the  (semimajor, semiminor) axes of the inner 
and outer ellipses, concentric to the bounding ellipse, and
constrained to lie in the angular sector that defines the elliptical
panda segment.  We use three rectangular regions to more precisely map 
the projected curved morphology of the outer tail. The centers in 
J2000 WCS coordinates, (length,width), and orientation angle of tail 
spectral regions are listed in Table \ref{tab:n7618tailregs}.
  
\begin{deluxetable}{cccc}
\tablewidth{0pc}
\tablecaption{Bounding Ellipse Parameters for UGC\,12491 Surface
    Brightness Profiles and Spectral Regions}\label{tab:ugc12491bounding}
\tablehead{\colhead{Region} & \colhead{Center} & \colhead{a,b}
  & \colhead{$\theta$} \\
  &(R.A.,Decl.)   & (arcsec, arcsec) & (deg)\\
($1$) & ($2$) & ($3$) & ($4$)}
\startdata
Nose  & $23^h18^m39.687^s,+42^\circ57^\prime15.366\as$ & 
      $56.1$, $50.195$ & $0$ \\ 
North Wing & $23^h18^m39.418^s,+42^{\circ}57^{\prime}18.319\as$ & 
      $56.172$, $50.259$ & $0$ \\
Inner W  & $23^{h}18^{m}38.2^{s},+42^{\circ}57^{\prime}29.0\as$ & 
      $21.55$,$12.1$ & $254.8$\\ 
Inner E Tail  & $23^{h}18^{m}38.2^{s},+42^{\circ}57^{\prime}29.0\as$ &
      $21.55$, $12.1$ & $254.8$ \\
Outer Tail N & $23^{h}18^{m}38.2^{s},+42^{\circ}57^{\prime}29.0\as$ &
      $21.55$,$12.1$ & $254.8$ \\ 
Outer Tail S & $23^{h}18^{m}38.2^{s},+42^{\circ}57^{\prime}29.0\as$ &
      $21.55$, $12.1$  & $254.8$\\
South Wing & $23^{h}18^{m}38.253^{s},+42^{\circ}57^{\prime}29.145\as$ &
      $65.1$, $28.494$ & $27$ \\
\enddata
\tablecomments{ Column ($2$) lists the J2000 WCS coordinates of the
  ellipse center,  column ($3$) lists the (semimajor, semiminor) axes of
  the bounding ellipse, and column ($4$) lists the position angle measured
  counterclockwise from west to the major axis of the ellipse. }
\end{deluxetable}

\begin{deluxetable}{ccccc}
\tablewidth{0pc}
\tablecaption{{UGC\,12491 Spectral Regions}\label{tab:ugc12491specregs}}
\tablehead{\colhead{Region} & \colhead{a$_i$,b$_i$} &
    \colhead{a$_o$,b$_o$} & \colhead{A} & \colhead{B} \\ 
    & (arcsec, arcsec) & (arcsec, arcsec) & (deg) & (deg) \\
    ($1$) & ($2$) & ($3$) & ($4$)& ($5$) }
\startdata
Nose1        & $33.474,29.950$  & $44.327,39.661$ & $333$ & $38.2$  \\
Nose2        & $44.327,39.661$  & $56.100,50.195$ & $333$ & $38.2$  \\
Nose3        & $56.100,50.195$  & $87.535,78.321$ & $333$ & $38.2$  \\
Nose4        & $87.535,78.321$  & $129.865,116.195$ & $333$ & $38.2$  \\
Nose5        & $129.865,116.195$ & $176.258,157.704$ & $333$ & $38.2$  \\
Nose6         &$176.258,157.704$ & $216,193.264$  & $333$ & $38.2$  \\
Nose7         &$216,193.264$    &$360,322.107$  & $333$ & $38.2$  \\
North Wing1   & $33.474,29.951$ & $56.172,50.259$ & $57.2$  & $96.75$ \\  
North Wing2   & $56.172,50.259$ & $78.782,70.489$ & $57.2$  & $96.75$ \\
North Wing3   & $78.782,70.489$ & $131.877,117.996$ & $57.2$  & $96.75$ \\
North Wing4  & $131.877,117.996$ & $200.861,179.718$ & $57.2$  & $96.75$ \\
North Wing5   &$200.861,179.718$&$290.000,259.474$& $57.2$  & $96.75$ \\
Inner W1      & $9.0,5.055$ & $29.60,16.625$  & $276$ & $22.4$ \\
Inner W2      & $29.60,16.625$ & $48.031,26.976$ & $276$ & $22.4$ \\
Inner Tail E1 & $9.0,5.055$   & $24.654,13.846$  & $120.4$  & $208.5$ \\
Inner Tail E2 & $24.654,13.846$ & $52.636,29.562$  & $120.4$  & $208.5$ \\
Inner Tail E3 & $52.636,29.562$ &$90.925,51.067$  & $120.4$  & $208.5$ \\
Inner Tail E4 & $90.925,51.067$ &$139.483,78.339$ & $120.4$  & $208.5$ \\
Outer Tail N  & $139.483,78.339$ & $239.284,134.391$ & $108.4$  & $141.3$   \\
Outer Tail S  & $139.483,78.339$ & $239.284,134.391$ & $141.3$  & $184.8$   \\
South Wing1   & $65.100,28.494$ & $92.44,40.461$ & $219.2$ & $305.8$ \\
South Wing2   & $92.44,40.461$ & $135.540,59.326$ & $219.2$&$305.8$ \\
South IGM$^a$ & $205,115.3$ &\ldots & \ldots & \ldots \\
\enddata
\tablecomments{Definitions of the spectral regions shown in the lower 
panel of Figure \protect\ref{fig:ugcanalysis}. 
For each elliptical panda segment, column ($2$) (a$_i$, b$_i$) and 
column ($3$) (a$_o$, b$_o$) give the inner and outer
(semimajor, semiminor) axes of the inner and outer ellipses,
respectively, concentric to the bounding ellipse given in Table
\protect\ref{tab:ugc12491bounding}. The spectral region is constrained
to lie in
the sector measured from angle A (column ($4$)) to angle B (column
($5$)). All angles are measured counterclockwise from west.   
{\em a} South IGM is a rectangular region centered at J2000 WCS coordinates
$\alpha=23^{h}18^{m}56.84^{s},\delta=+42^{\circ}54^{'}51.75\as$ with
(length, width) in column ($2$) and orientation angle $2^\circ.32$ measured
counterclockwise from west to the length axis.} 
\end{deluxetable}

\begin{deluxetable}{ccc}
\tablewidth{0pc}
\tablecaption{{Bounding Ellipse Parameters for NGC\,7618 Surface
    Brightness Profiles and Spectral Regions}\label{tab:n7618bounding}}
\tablehead{\colhead{Region} & \colhead{a,b} & \colhead{$\theta$} \\
  & (arcsec, arcsec) & (deg) \\
($1$) & ($2$) & ($3$) }
\startdata
Head          &$60.496,46.996$ &$51$  \\
North Plume   &$96.51,46.0$ & $44.64$  \\
West Wing     &$94.367,53.762$ &$309.65$  \\
East  Wing    &$96.510,46.00$ &$44.64$  \\
South Tail    &$86.388,43.194$ & $0$ \\
\enddata
\tablecomments{Each bounding ellipse is centered at WCS coordinates $\alpha =
  23^{h}19^{m}47.22^{s}$,$\delta=+42^{\circ}51^{'}09.5\as$, coincident
  with the X-ray peak (and optical center) of NGC\,7618. Column ($2$) lists the 
  (semimajor, semiminor) axes of the bounding ellipse. Column ($3$) 
  lists the position angle measured
  counterclockwise from west to the major axis of the ellipse. } 
\end{deluxetable}

\begin{deluxetable}{ccccc}
\tablewidth{0pc}
\tablecaption{{NGC\,7618 Elliptical Panda Spectral
    Regions}\label{tab:n7618specregs}}
\tablehead{\colhead{Region} & \colhead{a$_i$,b$_i$} &
    \colhead{a$_o$,b$_o$} & \colhead{A} & \colhead{B} \\ 
    & (arcsec, arcsec) & (arcsec, arcsec) & (deg) & (deg)\\
($1$) & ($2$) & ($3$) & ($4$) & ($5$)}
\startdata
Head1 &$5.0,3.884$  &$60.5,47.0$ &$53.4$ &$146.4$  \\
Head2 &$60.5,47.0$  &$200.0,155.372$ &$53.4$ &$146.4$  \\
Plume1 &$5.0,2.383$ &$96.51,46.0$ &$85.7$ &$155.8$  \\
Plume2 &$96.51,46.0$  &$213.251,101.643$ &$85.7$ &$155.8$  \\
Plume3 &$213.251,101.643$ &$396.040,188.8$ &$85.7$ &$155.8$  \\ 
East Wing1 &$5.0,2.383$  &$96.51,46.0$ &$155.9$ & $200.0$ \\ 
East Wing2 &$96.51,46.0$  &$348.1,165.9$ &$155.9$ & $200.0$ \\
West Wing1 &$4.958,2.825$  &$94.367,53.762$ &$342.65$ &$39$ \\ 
West Wing2 &$94.367,53.762$ &$179.09,102.03$ &$342.65$ &$39$ \\
West Wing3 &$179.09,102.03$ &$475.0,270.614$ &$342.65$ &$39$ \\
South Tail1 &$10.0,5.0$  &$86.388,43.194$ &$220$ & $318$ \\
South Tail2  &$86.388,43.194$  &$300.0,150.0$ &$220$ & $318$ \\
\enddata
\tablecomments{Definitions of the spectral regions shown in the lower 
panel of Figure \protect\ref{fig:ngcanalysis}. 
Column ($2$) (a$_i$, b$_i$) and column ($3$) (a$_o$, b$_o$) give
the inner and outer
(semimajor, semiminor) axes of the inner and outer ellipses,
respectively, for each elliptical panda segment concentric to
the bounding ellipse given in Table
\protect\ref{tab:n7618bounding}. The spectral region is constrained to lie in
the sector measured from angle A (column ($4$)) to angle B
(column ($5$)). All angles are measured counterclockwise from the west. }
\end{deluxetable}

\begin{deluxetable}{cccc}
\tablewidth{0pc}
\tablecaption{{NGC\,7618 Rectangular Tail  Spectral
    Regions}\label{tab:n7618tailregs}}
\tablehead{\colhead{Region} & \colhead{Center} & \colhead{l,w}
  & \colhead{$\theta$} \\
  & (R.A.,Decl.) & (arcsec,arcsec) & (deg) \\
($1$) & ($2$) & ($3$) & ($4$)}
\startdata
Southern Most Tail &
  $23^{h}19^{m}50.598^{s},+42^{\circ}48^{'}06.984\as$ & 
      $183.5,55.3$  & $359.9$ \\
Mid Tail &$23^{h}20^{m}03.417^{s},+42^{\circ}50^{'}21.839\as$ 
  &$150.584,115.63$ &$132.4$ \\
North Tail &$23^{h}20^{m}10.733^{s},+42^{\circ}52^{'}18.688\as$ &
              $123.16,100.4$&$310.3$\\
\enddata
\tablecomments{Rectangular spectral regions shown in 
  Figure \protect\ref{fig:ngcanalysis} used to measure the properties of
  hot gas in the outer tail regions of  NGC\,7618. Column ($2$) lists
  the center of the rectangular regionin J000 WCS coordinates, 
  column ($3$) lists the (length,width) of the region, and column ($4$) is
  the orientation angle of the rectangular region measured
  counterclockwise from west to the length axis. }  
\end{deluxetable}
\end{appendix}
\end{document}